\documentclass[twocolumn,preprintnumbers,amsmath,amssymb]{revtex4}

\usepackage{graphicx}
\usepackage{dcolumn}
\usepackage{bm}
\usepackage{color}
\usepackage{txfonts}
\def\i{i}
\def\d{d}
\def\e{e}
\def\vector#1{{\boldsymbol{#1}}}

\def\vd{{\vector d}}
\def\vg{{\vector g}}

\def\vk{{\vector k}}

\def\vq{{\vector q}}

\def\vR{{\vector R}}
\def\vS{{\vector S}}

\def\vS{{\vector S}}

\def\Tc{{T_{\rm c}}}

\def\dps{\displaystyle}

\def\kB{{k_{\rm B}}}

\def\hsp#1{\hspace{#1ex}}

\def\lsim{\stackrel{{\textstyle<}}{\raisebox{-.75ex}{$\sim$}}}
\def\gsim{\stackrel{{\textstyle>}}{\raisebox{-.75ex}{$\sim$}}}

\def\eq.#1{eq.~(\ref{#1})}
\def\eqs.#1{eqs.~(\ref{#1})}

\def\refeq#1{(\ref{#1})}

\newcommand\Equation[2]{
\begin{equation}\label{#1}
#2
\end{equation}
}

\begin{document}
\preprint{APS/123-QED}

\title{
Phase diagrams of noncentrosymmetric superconductors 
}


\author{Hiroshi Shimahara}



\affiliation{
Department of Quantum Matter Science, ADSM, Hiroshima University, 
Higashi-Hiroshima 739-8530, Japan
}


\date{Received November 15, 2012}

\begin{abstract}
Noncentrosymmetric superconductors 
with various types of pairing interactions 
are systematically examined 
with particular focus on phenomena that originate from 
the differences between Fermi surfaces split by a strong spin-orbit coupling. 
In particular, 
when the spin-orbit coupling increases and 
one of the split Fermi surfaces disappears, 
the phase diagram and the structure of the gap function 
change drastically. 
For example, we examine the conditions for the transition 
from full-gap states to line-node states (FLT), 
which may explain the differences in the experimental results 
between the noncentrosymmetric superconductors 
${\rm Li_2Pd_3B}$ and ${\rm Li_2Pt_3B}$ discovered recently. 
The dominant pairing interactions and gap functions 
can be predicted to some extent by comparing the theoretical and 
experimental results for these compounds. 
For example, if the FLT occurs by replacing Pd with Pt, 
it is most likely that the superconductivity is mainly induced by 
charge-charge interactions, 
and if this is the case, 
the superconductivities in ${\rm Li_2Pd_3B}$ and ${\rm Li_2Pt_3B}$ 
are an \mbox{s-wave} nearly spin-triplet state 
and a \mbox{d-wave} state that has both spin-singlet and triplet components of 
comparable weights, respectively. 
Comparing the theoretical phase diagrams in simple models, 
it is found that 
the FLT occurs in a wider realistic parameter region 
for charge-charge interactions, {\it i.e.}, 
where short-range Coulomb repulsion is strong and 
\mbox{p-wave} and \mbox{d-wave} interactions are attractive, 
while it occurs in narrower rather unrealistic parameter regions 
for interactions of magnetic origin. 
It is also found that \mbox{d-wave} spin-triplet pairing may occur, 
when pairing interactions are of magnetic origin and 
anisotropic in spin space. 
\end{abstract}


\maketitle


\section{\label{sec:introduction}
Introduction 
}

Recently, superconductors without inversion symmetry 
have been studied extensively 
owing to their unconventional features~\cite{Gor01,Ser04,Fri04,
Tog04,Bad05,Nis05,Nis07,Yua06,Hay06,Fuj05,Fuj07,Yan07,Sam08,
LuY08,Haf09,Pee11, Shis11}. 
A strong spin-orbit coupling results in the splitting of electronic bands, 
in which the direction of the electron spin depends on momentum. 
As a result, Cooper pairs are not purely spin-singlet or spin-triplet. 
Furthermore, interband pairing is forbidden, 
when spin-orbit coupling is so strong that 
the energy difference of spin-orbit split bands is larger than 
the magnitude of the gap function.

We are interested in the ternary borides 
${\rm Li_2Pd_3B}$ and ${\rm Li_2Pt_3B}$~\cite{Tog04,Bad05} 
among noncentrosymmetric superconductors, 
because their superconductivities exhibit completely different behaviors 
in spite of their same crystal structure. 
In nuclear magnetic resonance (NMR) measurement of ${\rm Li_2Pd_3B}$, 
Nishiyama {\it et al.} observed 
that the nuclear spin relaxation rate $T_1^{-1}$ 
exhibited a coherence peak just below $\Tc$, 
and the spin susceptibility decreased below $\Tc$~\cite{Nis05}. 
These results indicate that the gap function is isotropic 
and has components of antiparallel spin pairing. 
On the other hand, in ${\rm Li_2Pt_3B}$, 
the relaxation rate $T_1^{-1}$ did not exhibit any coherence peak 
and was proportional to $T^3$ below $\Tc$~\cite{Nis07}. 
These behaviors indicate that the gap function has line nodes. 
The low-temperature penetration depth $\lambda(T)$ 
measured by Yuan {\it et al.} exhibited a BCS-like behavior 
in ${\rm Li_2Pd_3B}$, 
while it exhibited a linear temperature dependence in ${\rm Li_2Pt_3B}$, 
which also supports the existence of line nodes~\cite{Yua06}.

In ${\rm Li_2Pt_3B}$, 
the Knight shift remained unchanged across $\Tc$~\cite{Nis07} 
in contrast to that in ${\rm Li_2Pd_3B}$. 
The theoretical explanation for this behavior seems difficult 
because of the following. 
If the behavior indicates that the spin susceptibility remains unchanged 
across $\Tc$, 
antiparallel-spin pairing is excluded. 
On the other hand, as Frigeri {\it et al.} have shown~\cite{Fri04}, 
the \mbox{d-vector} $\vd$ must be parallel to 
the direction of the momentum-dependent spin axis ${\hat \vg}(\vk)$ 
in noncentrosymmetric superconductors with a strong spin-orbit coupling. 
Below, we shall argue that these results lead to a contradiction, 
unless there is any extra effect considered.

The results of specific heat measurement and muon-spin rotation experiment 
by H${\rm \ddot a}$fliger {\it et al.} indicate that 
the whole family of ${\rm Li}_2({\rm Pd}_{1-x}{\rm Pt}_x)_3{\rm B}$ 
comprises single-gap \mbox{s-wave} superconductors 
across the entire doping regime~\cite{Haf09}. 
The $H$-$T$ phase diagram and several superconducting parameters 
obtained by Peets {\it et al.} exhibit a continuous change 
as functions of the doping ratio $x$~\cite{Pee11}. 
Therefore, 
the pairing symmetries of these compounds are still controversial.

Recently, Shishidou and Oguchi have performed 
first-principles calculation 
in ${\rm Li_2Pd_3B}$ and ${\rm Li_2Pt_3B}$ 
and obtained Fermi surface structures~\cite{Shis11}. 
A strong spin-orbit coupling results in a large splitting of Fermi surfaces. 
In each of the spin-orbit split bands, 
the direction of the electron spin depends on momentum. 
According to their results, 
every Fermi surface appears to have their partners of 
spin-orbit split Fermi surfaces (SFSs) in ${\rm Li_2Pd_3B}$, 
while some of the Fermi surfaces do not appear to have their partners 
in ${\rm Li_2Pt_3B}$ owing to the stronger spin-orbit coupling, 
although strictly speaking the relations of spins and momenta 
on the SFSs are quite complicated.

In this study, motivated by the above experimental and theoretical results, 
we examine the phase diagrams of pairing anisotropy 
in systems with a strong spin-orbit coupling. 
In particular, we focus on possible drastic changes in the superconductivity 
when one of the SFSs disappears. 
For example, the experimental and theoretical results mentioned above 
seem to suggest that a full-gap state changes into a line-node state 
when the spin-orbit coupling increases and one of the SFSs disappears. 
We abbreviate such a full-gap line-node transition as FLT hereafter. 
Such a behavior may be attributed 
both to the changes in the electron states and 
to those in the phonon states. 
We examine the former possibility in this study. 
Although we call such a change a transition, 
it is not necessarily a phase transition that exhibits a discontinuity 
at a specific spin-orbit coupling constant. 
In real materials, with increasing coupling constant, 
the density of states from Fermi surfaces without 
spin-orbit split partners may increase continuously. 
In this case, averaged physical quantities contributed by 
both kinds of Fermi-surfaces with and without partners 
may change continuously.

In~\S~\ref{sec:form}, we briefly review the formulation used in this study. 
Possible forms of gap functions are shown, 
and Frigeri {\it et al.}'s result mentioned above is reproduced. 
In~\S~\ref{sec:super}, we derive the expressions of the dimensionless 
coupling constants and the transition temperatures of the superconductivity 
on the basis of a model with intraband pairing interactions and 
interband pair-hopping interactions. 
We pay special attention to the differences between the two SFSs. 
In~\S~\ref{sec:pairinginteraction}, 
we derive intraband pairing interactions 
and interband pair-hopping interactions 
from original interactions between electrons with momentum-independent spins. 
We suppose the charge-charge interaction (CI) 
and the spin-spin interaction (SI) 
as original interactions. 
In~\S~\ref{sec:limits}, we examine two limiting cases, 
{\it i.e.}, an equal-band limit and a single-band limit. 
The latter case occurs when one of the SFSs disappears 
owing to a stronger spin-orbit coupling. 
In~\S~\ref{sec:phasediagrams}, in order to illustrate our theory, 
we examine spherically symmetric systems as examples. 
Phase diagrams in planes of the coupling constants are shown 
for several types of interactions. 
In~\S~\ref{sec:last}, we summarize the results and 
discuss ternary superconductors. 
We use the units where $\hbar = 1$ and $\kB = 1$.

\section{\label{sec:form}
Formulation 
}

First, we examine the Hamiltonian of noninteracting electrons defined by 
\Equation{eq:H0}
{
     H_0 = \sum_{\vk} 
             \left ( 
               \begin{array}{cc} 
                 c_{\vk \uparrow}^{\dagger}, & 
                 c_{\vk \downarrow}^{\dagger} 
               \end{array}
             \right ) 
             \left ( 
               {\hat \epsilon}_{\vk} - \mu \sigma_0 
             \right ) 
             \left ( 
               \begin{array}{c} 
                 c_{\vk \uparrow} \\ 
                 c_{\vk \downarrow} 
               \end{array}
             \right ) , 
     }
with 
\Equation{eq:epsilon}
{
     {\hat \epsilon}_{\vk} 
       \, = \, 
         \epsilon_{\vk}^0 \, \sigma_0 
         - \alpha \, {\hat {\vg}}(\vk) \cdot \boldsymbol{\sigma} , 
     }
where $\sigma_0$ and $\boldsymbol{\sigma}$ are 
the $2 \times 2$ identity matrix and Pauli matrix, respectively. 
We suppose the vector function ${\hat \vg}(\vk)$ that satisfies 
${\hat \vg}(-\vk) = - {\hat \vg}(\vk)$ and \mbox{$|{\hat \vg}(\vk)| = 1$}, 
and express it as 
\Equation{eq:gthetavarphidef}
{
     \begin{array}{rcl} 
     {\hat \vg}(\vk) & = & (g_x(\vk), g_y(\vk), g_z(\vk)) \\
         & = & (\sin {\bar \theta}_{\vk} \cos {\bar \varphi}_{\vk}, \\ 
         &&  ~~ \sin {\bar \theta}_{\vk} \sin {\bar \varphi}_{\vk}, 
                            \cos {\bar \theta}_{\vk} ) , 
     \end{array}
     }
with the polar coordinates $({\bar \theta}_{\vk},{\bar \varphi}_{\vk})$. 
We divide the momentum space into two regions $R_{\pm}$, such that 
$$
     \vk \in R_{\pm} ~~ \Leftrightarrow ~~ \pm g_y(\vk) > 0 , 
     $$
and define unitary matrices by 
$${
     \begin{array}{rcl}
     U_{\vk} & = & R_z({\bar \varphi}_{\vk}) \, R_y({\bar \theta}_{\vk}) \\[8pt]
     & = & \dps{ 
     \left ( 
       \begin{array}{cc}
       {   \e^{ - \i \frac{{\bar \varphi}_{\vk}}{2} } 
             \cos \frac{{\bar \theta}_{\vk}}{2} } & 
       { - \e^{ - \i \frac{{\bar \varphi}_{\vk}}{2} } 
             \sin \frac{{\bar \theta}_{\vk}}{2} } \\[8pt]
       {   \e^{   \i \frac{{\bar \varphi}_{\vk}}{2} } 
             \sin \frac{{\bar \theta}_{\vk}}{2} } & 
       {   \e^{   \i \frac{{\bar \varphi}_{\vk}}{2} } 
             \cos \frac{{\bar \theta}_{\vk}}{2} } 
       \end{array}
       \right ) 
     }
     \end{array}
     }$$
$$ {
     \begin{array}{rcl}
     U_{-\vk} & = & - \i    R_z({\bar \varphi}_{-\vk}) 
                         \, R_y({\bar \theta}_{-\vk})  \\[8pt]
     & = & \dps{ 
     \left ( 
       \begin{array}{cc}
       { - \e^{ - \i \frac{{\bar \varphi}_{\vk}}{2} } 
             \sin \frac{{\bar \theta}_{\vk}}{2} } & 
       {   \e^{ - \i \frac{{\bar \varphi}_{\vk}}{2} } 
             \cos \frac{{\bar \theta}_{\vk}}{2} } \\[8pt]
       {   \e^{   \i \frac{{\bar \varphi}_{\vk}}{2} } 
             \cos \frac{{\bar \theta}_{\vk}}{2} } & 
       {   \e^{   \i \frac{{\bar \varphi}_{\vk}}{2} } 
             \sin \frac{{\bar \theta}_{\vk}}{2} } 
       \end{array}
       \right ) 
     }
     \end{array}
     }$$ 
for $\vk \in R_{+}$. We transform the electron operators $c_{\vk \sigma}$ 
into fermion operators ${\tilde c}_{\vk \pm}$ by 
$({\tilde c}_{\vk +}^{\dagger}, {\tilde c}_{\vk -}^{\dagger}) 
 = (c_{\vk \uparrow}^{\dagger}, c_{\vk \downarrow}^{\dagger}) U_{\vk}$. 
These transformations are essentially the same as those used 
in previous studies~\cite{Gor01,Ser04,Sam08}. 
Using $U_{\vk}$ and $U_{-\vk}$, the Hamiltonian $H_0$ is diagonalized as 
$${
     H_0 = \sum_{s=\pm} \, 
       \sum_{ \vk \in R_{+}} 
       {\tilde \xi}_{\vk s} 
       \Bigl (   {\tilde c}_{\vk s}^{\dagger} {\tilde c}_{\vk s} 
               + {\tilde c}_{- \vk s}^{\dagger} {\tilde c}_{- \vk s} \Bigr ) 
     }$$ 
with ${\tilde \xi}_{\vk s} = \epsilon_{\vk}^{0} - s \alpha - \mu$.

Next, we examine the Cooper-pair operators defined by 
${\hat \psi}_{\sigma \sigma'} (\vk) 
       \equiv c_{\vk\sigma} c_{- \vk \sigma'}$ 
and 
${\tilde \psi}_{s s'} (\vk) 
       \equiv {\tilde c}_{\vk s} {\tilde c}_{- \vk s'}$. 
In terms of the \mbox{d-vector} 
${\hat \vd}(\vk) 
= \bigl ({\hat d}_x(\vk), {\hat d}_y(\vk), {\hat d}_z(\vk) 
  \bigr )$, 
and the singlet component $d_0(\vk)$, 
the Cooper-pair operators are expressed as 
$$ {
     \begin{array}{l}
     \left ( 
     \begin{array}{cc}
     {\hat \psi}_{\uparrow \uparrow}(\vk) & 
     {\hat \psi}_{\uparrow \downarrow}(\vk) \\
     {\hat \psi}_{\downarrow \uparrow}(\vk) & 
     {\hat \psi}_{\downarrow \downarrow}(\vk) 
     \end{array}
     \right ) 
     \\[12pt]
     ~~~ \equiv 
     \left ( 
     \begin{array}{cc}
     - {\hat d}_x(\vk) + \i {\hat d}_y(\vk) & 
       {\hat d}_z(\vk) + {\hat d}_0(\vk) \\ 
       {\hat d}_z(\vk) - {\hat d}_0(\vk) & 
       {\hat d}_x(\vk) + \i {\hat d}_y(\vk) \\ 
     \end{array}
     \right ) . 
     \end{array}
     }$$ 
The unitary transformations defined above lead to 
\Equation{eq:psitransformed}
{
     \begin{array}{rcl} 
     {\tilde \psi}_{++}(\vk) 
       & = & 
         s_{\vk} \, 
         \bigl ( {\hat \vg}(\vk) \cdot {\hat \vd}(\vk) + {\hat d}_0(\vk) \bigr ) 
         \\
     {\tilde \psi}_{--}(\vk) 
       & = & 
         s_{\vk} \, 
         \bigl( {\hat \vg}(\vk) \cdot {\hat \vd}(\vk) - {\hat d}_0(\vk) \bigr ) 
         \\
     {\tilde \psi}_{+-}(\vk) 
       & = & 
         {\vg}_{+-}(\vk) \cdot {\hat \vd}(\vk) 
         \\
     {\tilde \psi}_{-+}(\vk) 
       & = & 
         {\vg}_{-+}(\vk) \cdot {\hat \vd}(\vk) , 
     \end{array}
     }
with $s_{\vk} = \pm 1$ for $\vk \in R_{\pm}$, 
where we have introduced the vectors 
\Equation{eq:gsdef}
{
     \begin{array}{rcl} 
     \vg_{+-}(\vk) & \equiv & 
       \bigl ( - \cos {\bar \theta}_{\vk} \cos {\bar \varphi}_{\vk} 
               - \i \sin {\bar \varphi}_{\vk} , \\ 
     && ~~~~~ 
             ~ - \cos {\bar \theta}_{\vk} \sin {\bar \varphi}_{\vk} 
               + \i \cos {\bar \varphi}_{\vk} , 
             ~   \sin {\bar \theta}_{\vk} \bigr ) , 
     \end{array}
     }
and $\vg_{-+}(\vk) = - \vg_{+-}^{*}(\vk) $. 
All three vectors ${\vg}_{\pm \mp}(\vk)$ and ${\hat \vg}(\vk)$ 
are orthogonal to each other.

When $\alpha \gg \kB \Tc$, 
we have $\langle {\tilde \psi}_{\pm \mp} (\vk) \rangle = 0$ for any $\vk$. 
This condition, 
together with \eqs.{eq:psitransformed} and \refeq{eq:gsdef}, 
immediately results in 
$\langle {\hat \vd}(\vk) \rangle \parallel {\hat \vg}(\vk)$, 
which coincides with 
the result obtained by \mbox{Frigeri \it et al.}~\cite{Fri04} 
Hence, we can define the scalar operator ${\hat d}(\vk)$ such that 
${\hat \vd}(\vk) = {\hat d}(\vk) {\hat \vg}(\vk)$. 
Since $\vd(\vk)$ and ${\hat \vg}(\vk)$ are of odd parity, 
the operator ${\hat d}(\vk)$ is of even parity. 
In terms of ${\hat d}(\vk)$ and ${\hat d}_0(\vk)$, 
the Cooper-pair operators are rewritten as 
\Equation{eq:psilargealpha}
{
     {\tilde \psi}_{ss}(\vk) 
       = s_{\vk} \, 
         \bigl ( {\hat d}(\vk) + s {\hat d}_0(\vk) \bigr ) . 
     }
The results of this section do not depend on the form of 
pairing interactions.

\section{\label{sec:super}
Superconductivity 
}

In the weak-coupling theory, the pairing interactions are expressed by 
\Equation{eq:H1def}
{
     H_1 = \frac{1}{N} \! 
       \sum_{\vk,\vk'}
       \sum_{\mu = 0}^{3} 
         V_{\mu}(\vk,\vk') \, {\hat d}_{\mu}^{\dagger}(\vk) \, 
                              {\hat d}_{\mu}(\vk') , 
     }
where we have neglected corrections due to the broken inversion symmetry. 
When $\alpha \gg \kB \Tc$, 
we can omit terms that include ${\tilde \psi}_{\pm \mp}$. 
Hence, \eq.{eq:H1def} is rewritten as 
\Equation{eq:H1_in_psi_ss}
{
     H_1 = \frac{1}{N} \! 
       \sum_{\vk,\vk'\in R_{+}} 
       \sum_{ ss' }
         \Gamma_{ss'}(\vk, \vk') \, 
           {\tilde \psi}_{ss}^{\dagger}(\vk) 
           {\tilde \psi}_{s's'}(\vk') , 
     } 
where 
\Equation{eq:GammaVV}
{
     \Gamma_{ss'}(\vk,\vk') 
       = ss' V_{\rm sin} (\vk,\vk') 
           + {\tilde V}_{\rm tri} (\vk,\vk') 
     }
for $\vk,\vk' \in R_{+}$, and 
\Equation{eq:VsingVtripdef}
{
     \begin{array}{rcl} 
     \dps{ 
       V_{\rm sin} (\vk,\vk') } & = & \dps{ V_0(\vk,\vk') } \\[8pt] 
     \dps{ 
       {\tilde V}_{\rm tri} (\vk,\vk') } & = & \dps{ 
         \sum_{\mu=1}^{3} 
           {\hat g}_{\mu}(\vk) V_{\mu}(\vk,\vk') {\hat g}_{\mu}(\vk') . 
           }\\
     \end{array}
     }
We define the gap function as 
\Equation{eq:Delta}
{
     \Delta_{\vk s} 
       = - \frac{1}{N} \sum_{\vk' \in R_{+}} \sum_{s'=\pm} 
             \Gamma_{s s'} (\vk,\vk') 
               \langle {\tilde \psi}_{s's'}(\vk') \rangle 
     }
and the temperature Green's functions as 
$${
     \begin{array}{rcl} 
     {\cal G}_{s}(\vk,\tau) 
     & = & \dps{ 
       - \langle T_{\tau} {\tilde c}_{\vk s}(\tau) 
                     {\tilde c}_{\vk s}^{\dagger} \rangle } \\[4pt] 
     {\cal F}_{s}(\vk,\tau) 
     & = & \dps{ 
       - \langle T_{\tau} {\tilde c}_{- \vk s}^{\dagger}(\tau)
                     {\tilde c}_{\vk s}^{\dagger} \rangle } , 
     \end{array}
     }$$
with $A(\tau) = \e^{\tau H} A \e^{- \tau H}$ and $H = H_0 + H_1$. 
The gap function is written as 
$${
     \Delta_{\vk s}^{*} 
       = \frac{1}{N} \sum_{\vk' \in R_{+}} \sum_{s'=\pm} 
             \Gamma_{s s'} (\vk,\vk') 
             {\cal F}_{s'}(\vk',-0) . 
     }$$
We obtain 
\Equation{eq:GFresults}
{
     \begin{array}{rcl} 
     {\cal G}_{s}(\vk,\omega_n) 
     & = & \dps{ 
       \frac{ \i \omega_n + {\tilde \xi}_{\vk s} }
            {( \i \omega_n - E_{\vk s} ) ( \i \omega_n + E_{\vk s} ) }
                     } \\[12pt] 
     {\cal F}_{s}(\vk,\omega_n) 
     & = & \dps{ 
       \frac{ \Delta_{\vk s}^{*} }
            {( \i \omega_n - E_{\vk s} ) ( \i \omega_n + E_{\vk s} ) } , 
                     } 
     \end{array}
     }
with the quasi-particle energy 
\Equation{eq:Eresult}
{
     E_{\vk s} = \sqrt{ {\tilde \xi}_{\vk s}^2 + | \Delta_{\vk s} |^2 } , 
     }
as previous authors have obtained~\cite{Gor01,Ser04,Fuj05,Sam08}. 
We obtain the self-consistent equation 
\Equation{eq:SCE}
{
     \Delta_{\vk s}
       = - \frac{1}{N} \sum_{\vk' \in R_{+}} \sum_{s'=\pm} 
             \Gamma_{s s'} (\vk,\vk') W(E_{\vk' s'})
             \Delta_{\vk' s'} , 
     }
where $W(E) = \tanh(E/2T)/2E$.

We assume that pairing interactions exist only between electrons 
near Fermi surfaces, 
when such interactions are mediated not only by phonons, 
but also by spin and charge fluctuations~\cite{Sam08,Shi03,Fay77}. 
This can be taken into account 
by introducing effective cutoff energies 
for each vertex function $\Gamma_{ss'}(\vk,\vk')$. 
In general, the cutoff energy depends on the positions of 
the interacting electrons on the Fermi surfaces. 
In particular, 
we retain the dependence on the band indexes of the interacting electrons. 
Therefore, the gap functions are written in the form 
\Equation{eq:Deltacutoff}
{
     \Delta_{\vk s} 
       = \Delta_{{\hat \vk}}^{(s)} 
           \theta( \omega_{\rm c}^{(s)} - | {\tilde \xi}_{\vk s} | ) , 
     }
where ${\hat \vk} = \vk/|\vk|$, 
and the pairing interactions are written in the separable forms 
\Equation{eq:Gammacutoff}
{
     \Gamma_{ss'}(\vk,\vk') 
       = \Gamma_{{\hat \vk}{\hat \vk'}}^{(ss')} 
           \theta( \omega_{\rm c}^{(s)}  - | {\tilde \xi}_{\vk s}   | ) 
           \theta( \omega_{\rm c}^{(s')} - | {\tilde \xi}_{\vk' s'} | ) . 
     }
For the pairing interaction mediated by phonons, 
the cutoff frequencies $\omega_{\rm c}^{(s)}$ 
can be replaced with the Debye frequency $\omega_{\rm D}$, 
which does not strongly depend on the band index $s = \pm$. 
For those mediated by electronic fluctuations, 
they are characteristic energy scales of such fluctuations, 
which strongly depend on the band index $s = \pm$, 
because the nesting condition strongly depends on 
the shapes of the Fermi surfaces. 
The spin and charge susceptibilities have a sharp peak 
at the nesting vector $\vq_0$ 
that connects parts of the Fermi surfaces with a better nesting condition. 
Thus, pairing interactions mediated by corresponding fluctuations 
become strong at $\vq_0$, 
within a momentum width comparable to the peak width of 
the corresponding susceptibility~\cite{Shi89}. 
Since the peak width reflects the difference 
between the original Fermi surface and that shifted by the nesting vector, 
the cutoff frequencies $\omega_{\rm c}^{(s)}$ are energy scales 
that correspond to the peak width in momentum space. 
For example, 
a smaller $\omega_{\rm c}^{(+)}$ or $\omega_{\rm c}^{(-)}$ means 
a critical slowing down of such fluctuations 
in proximity to the corresponding phase transition. 
Therefore, it is worth examining the effect of the difference between 
$\omega_{\rm c}^{(+)}$ and $\omega_{\rm c}^{(-)}$ on the superconductivity.

We rewrite the gap equation in the above model. 
By introducing the density of states $\rho_{s}$ defined by 
$${
     \frac{1}{N} \sum_{\vk} F({\tilde \xi}_{\vk s}, {\hat \vk}) 
     = 
       \int \frac{\d \Omega_{\hat \vk}}{4 \pi} 
       \int \d \xi \, 
         \rho_{s}(\xi,{\hat \vk}) 
         F(\xi, {\hat \vk}) , 
     }$$
where $F(\xi,{\hat \vk})$ is an arbitrary function, 
\eq.{eq:SCE} is written in the form 
$$ {
     \begin{array}{rcl} 
     \Delta_{{\hat \vk}}^{(s)} 
       & = & \dps{ 
         - \frac{1}{2} \sum_{s'=\pm} 
         \int \frac{\d \Omega_{{\hat \vk}'}}{4 \pi} 
             \rho_{s'}(0,{\hat \vk}') \, 
             \Gamma_{{\hat \vk}{\hat \vk}'}^{(ss')}
         } \\
       && \dps{ ~ \times 
         \int_{-\omega_{\rm c}^{(s')}}^{\omega_{\rm c}^{(s')}} \d \xi \, 
             W \bigl ( 
                 (\xi^2 + |\Delta_{{\hat \vk}'}^{(s')}|^2 )^{\frac{1}{2}} 
               \bigr ) 
             \Delta_{{\hat \vk}'}^{(s')} . 
         }
     \end{array} 
     }$$ 
By assuming the second-order phase transition, 
the superconducting transition temperature $\Tc$ is given by 
the condition of the first appearance of the nontrivial solution of 
the eigen equations 
\Equation{eq:Tceq}
{
     \begin{array}{rcl} 
     \Delta_{{\hat \vk}}^{(s)} 
       & = & \dps{ 
         - \frac{1}{2} \sum_{s'=\pm} 
         \int \frac{\d \Omega_{{\hat \vk}'}}{4 \pi} 
             \rho_{s'}(0,{\hat \vk}') \, 
             \Gamma_{{\hat \vk}{\hat \vk}'}^{(ss')}
         } \\
       && \dps{ ~ \times 
         \ln \frac{2 \e^{\gamma} \omega_{\rm c}^{(s')}}
                  {\pi \Tc}
             \Delta_{{\hat \vk}'}^{(s')} , 
         }
     \end{array} 
     }
where $\gamma = 0.57721 \cdots$ is Euler's constant.

On the basis of \eqs.{eq:Deltacutoff} and \refeq{eq:Gammacutoff}, 
we introduce the basis functions 
$${
     \gamma_{\alpha}^{(s)} (\vk) \, 
       = \, \theta( \omega_{\rm c}^{(s)} - | {\tilde \xi}_{\vk s} | ) 
         \, \, \gamma_{\alpha}({\hat \vk}) 
     }$$
that are normalized by 
$$ {
     \frac{1}{N} \sum_{\vk} 
       \bigl [ {\gamma_{\alpha}^{(s)}} (\vk) \bigr ]^{*} 
       \gamma_{\alpha'}^{(s)} (\vk) 
         = \delta_{\alpha \alpha'} . 
     }$$ 
Here, $\alpha$ and $\gamma_{\alpha}({\hat \vk})$ 
denote a symmetry index and the corresponding basis function 
with respect to the direction of $\vk$, respectively. 
By choosing a set of basis functions 
that are compatible with the symmetry of the system, 
the pairing interactions are expressed as 
\Equation{eq:Valpha}
{
     V_{\mu}(\vk,\vk') 
       = \sum_{\alpha} 
           \bigl [ \gamma_{\alpha}^{(s)}(\vk) \bigr ]^{*} 
           V_{\mu \alpha}^{(ss')} 
           \gamma_{\alpha}^{(s')}(\vk') , 
     }
and 
$${
     \Gamma_{ss'}(\vk,\vk') 
       = 
           s_{\vk} 
           s_{\vk'} 
         \!\! 
         \sum_{\alpha {\rm (even)}} 
         \!\! 
           \bigl [ \gamma_{\alpha}^{(s)} (\vk) \bigr ]^{*} 
             \Gamma_{\alpha}^{(ss')} \, 
           \gamma_{\alpha}^{(s')} (\vk') . 
     }$$
We should note that only $\alpha$'s of even parity appear 
in the expansion of $\Gamma_{ss'}$ 
from \eqs.{eq:GammaVV} and \refeq{eq:VsingVtripdef}. 
With ${\tilde V}_{\alpha}^{(ss')}$ defined by 
$$ {
     \begin{array}{rcl}
     {\tilde V}_{\rm tri}(\vk,\vk') 
       & = & \dps{ 
         \!\! 
         \sum_{\alpha {\rm (even)}} 
         \!\! 
           \bigl [ \gamma_{\alpha}^{(s)} (\vk) \bigr ]^{*} 
             {\tilde V}_{\alpha}^{(ss')} \, 
           \gamma_{\alpha}^{(s')} (\vk') , 
           } 
     \end{array}
     }$$ 
we obtain 
\Equation{eq:Gamma_alpha_sintri}
{
     \Gamma_{\alpha}^{(ss')}
       = ss' V_{0\alpha}^{(ss')} + {\tilde V}_{\alpha}^{(ss')} 
     }
from \eq.{eq:GammaVV}. 
The gap functions are expressed as 
\Equation{eq:Deltaexpand}
{
     \Delta_{\vk s} 
       = 
           s_{\vk} \!\! 
           \sum_{\alpha {\rm (even)}} \!\! 
           \Delta_{\alpha}^{(s)}
           \gamma_{\alpha}^{(s)}(\vk) . 
     }

The linearized gap equation \refeq{eq:Tceq} is decoupled into 
a set of equations 
$${
     \Delta_{\alpha}^{(s)} 
       = \sum_{s'=\pm} 
           \lambda_{\alpha}^{(ss')} 
           \ln \frac{2 \e^{\gamma} \omega_{\rm c}^{(s')}}
                    {\pi T_{c\alpha}}
           \Delta_{\alpha}^{(s')} , 
     }$$
with 
\Equation{eq:lambdaalpha}
{
     \lambda_{\alpha}^{(ss')}
     = - \frac{1}{2} \Gamma_{\alpha}^{(ss')} 
           \rho_{s'}^{(\alpha)}(0) 
     }
and 
$$ {
     \rho_{s'}^{(\alpha)}(0) 
     \equiv 
       \int \frac{\d \Omega_{{\hat \vk}'}}{4 \pi} 
         \rho_{s'}(0,\vk') 
         \bigl | \gamma_{\alpha}(\vk') \bigr |^{2} . 
     }$$ 
Here, $T_{c \alpha}$'s denote the transition temperatures 
when $\Delta_{\alpha'}^{(s)} \propto \delta_{\alpha' \alpha}$ 
is assumed. 
The physical transition temperature, 
below which the nontrivial solution $\Delta_{\vk s} \ne 0$ exists, 
is given by $\Tc = \max_{\alpha} T_{c \alpha}$

When we restrict ourselves to the symmetry index $\alpha$ 
that gives the highest $T_{c\alpha}$, 
we omit the index as 
$\Delta_{s} = \Delta_{\alpha}^{(s)}$, 
$l_{s} = \ln (2 \e^{\gamma} \omega_{\rm c}^{(s)}/\pi T_{c})$ 
and define intra- and inter-band coupling constants as 
$\lambda_{s} = \lambda_{\alpha}^{(ss)}$ and 
$\lambda_{s}' = \lambda_{\alpha}^{(s,-s)}$, respectively. 
Hence, the linearized gap equation is written as 
\Equation{eq:lingapeq}
{
     \left ( 
       \begin{array}{cc}
       1 - \lambda_{+}  l_{+} &   - \lambda_{+}' l_{-} \\ 
         - \lambda_{-}' l_{+} & 1 - \lambda_{-}  l_{-} 
       \end{array}
     \right ) 
     \left ( 
       \begin{array}{c}
       \Delta_{+} \\ 
       \Delta_{-} 
       \end{array}
     \right ) 
     = 
     \left ( 
       \begin{array}{c}
       0 \\
       0
       \end{array}
     \right ) . 
     }
We introduce the arbitrary energy scale $\omega_{\rm c}$ comparable to 
$\omega_{\rm c}^{(\pm)}$ and define 
$\delta_{\pm} \equiv \ln (\omega_{\rm c}^{(\pm)}/\omega_{\rm c})$, 
and $l \equiv \ln (\omega_{\rm c}/T)$. 
We obtain the expression of the transition temperature 
\Equation{eq:Tcresult}
{
     \Tc = \frac{2 \e^{\gamma}} {\pi} \omega_{\rm c}
       \e^{- \frac{1}{\Lambda}} 
     }
with the effective coupling constant 
\Equation{eq:Lambda}
{
     \Lambda = 
       \frac{1}{2} 
       \left [ 
       {\tilde \lambda}_{+} + {\tilde \lambda}_{-}
       \pm 
       \sqrt{ 
           ( {\tilde \lambda}_{+} - {\tilde \lambda}_{-})^{2} 
         + 4 {\tilde \lambda}'_{+} {\tilde \lambda}'_{-} } 
       \right ] , 
     }
with 
${\tilde \lambda}_{\pm} 
  = (\lambda_{\pm} - \delta_{\mp} d_{\lambda})/(1 - {\tilde \delta})$, 
${\tilde \lambda}'_{\pm} 
  = \lambda'_{\pm}/( 1 - {\tilde \delta} )$, 
and 
${\tilde \delta} = \lambda_{+} \delta_{+} + \lambda_{-} \delta_{-} 
- \delta_{+} \delta_{-} d_{\lambda}$, 
$d_{\lambda} = \lambda_{+} \lambda_{-} - \lambda'_{+} \lambda'_{-}$. 
In \eq.{eq:Lambda}, we take the sign that gives a larger $\Tc$ and 
satisfies the condition that $\Lambda > 0$ and $\Tc \ll \omega_{\rm c}^{(\pm)}$. 
If we set $\omega_{\rm c} = \omega_{\rm c}^{(+)} = \omega_{\rm c}^{(-)}$, 
\eq.{eq:Lambda} with a $+$ sign is 
reduced to the expression obtained by Samokhin and Mineev~\cite{Sam08}. 
Defining 
$ p = { 
       \bigl \{ 
           \sqrt{ 1 +{4 r_{\lambda}}/{\lambda_{+} \lambda_{-} q^2} } 
           - 1 
         \bigr \}/2 
       }$, 
$ q = { 
       \lambda_{-}^{-1} - \lambda_{+}^{-1} 
       + (1 - r_{\lambda} ) l_{\omega} 
       }$, 
$r_{\lambda} = \lambda'_{+} \lambda'_{-}/\lambda_{+} \lambda_{-}$, 
and $l_{\omega} = \ln (\omega_{\rm c}^{(+)}/\omega_{\rm c}^{(-)})$, 
we obtain the compact form 
$${
     \frac{1}{\Lambda} 
       = \frac{1}{1 - r_{\lambda}} 
         \Bigl ( \frac{1}{\lambda_{\mp}} \mp pq \Bigr ) 
     }$$
and $\Tc = (2 \e^{\gamma}/\pi) \omega_{\rm c}^{(\pm)} \exp(- 1/\Lambda)$.

\begin{figure}[hbtp]
\vspace{2ex} 
\vspace{2ex} 
\begin{center}
\includegraphics[width=8.0cm]{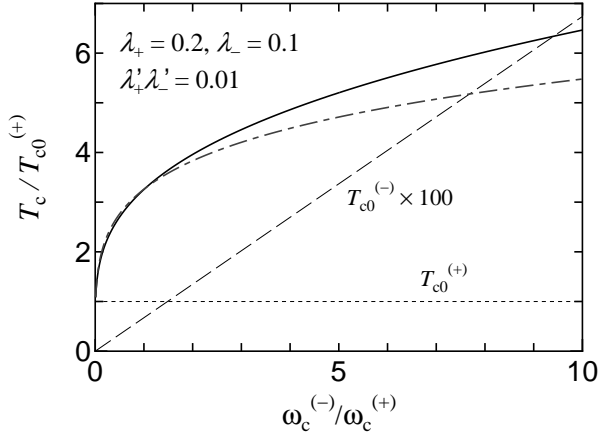}
\end{center}
\caption{Transition temperatures as functions of the cutoff energy. 
The solid and dashed curves show 
$\Tc$ and $T_{c0}^{(-)}$, respectively. 
As an example, 
$\lambda_{+} = 0.2$, $\lambda_{-}=0.1$, and 
$\lambda_{+}' \lambda_{-}' = 0.01$ are assumed. 
The dot-dash curve shows the result 
when the ratio $T_{c0}^{(-)}/T_{c0}^{(+)} = 0.007$ is fixed 
by adjusting $\lambda_{-}$. 
}
\label{fig:lomegadep}
\end{figure}

Figure~\ref{fig:lomegadep} shows 
the behaviors of the transition temperatures 
in the presence of the $\pm$ band mixing, 
$T_{c0}^{(s)}$ denotes the transition temperature of a single $s$--band 
with $\lambda_{\pm}' = 0$. 
Without losing generality, 
we have assumed $\lambda_{+} > \lambda_{-}$. 
Note that the scale of $T_{c0}^{(-)}$ is 
much smaller than those of $T_{c0}^{(+)}$ and $\Tc$ 
in Fig.~\ref{fig:lomegadep}. 
It is found that the presence of $\lambda_{-}$, 
even if it is so small that it gives a negligible $T_{c0}^{(-)}$, 
markedly enhances the transition temperature 
through the interband interactions $\lambda_{\pm}'$. 
The transition temperature increases 
as the ratio $\omega_{\rm c}^{(-)}/\omega_{\rm c}^{(+)}$ increases. 
We obtain essentially the same behavior 
when $T_{c0}^{(-)}$ is fixed by adjusting $\lambda_{-}$, 
as shown by the dot-dashed curve in Fig.~\ref{fig:lomegadep}. 
Therefore, the imbalance in $\omega_{\rm c}^{(\pm)}$ tends to enhance the 
transition temperature through the interband mixing effect. 
As argued above, 
the model with $\omega_{\rm c}^{(-)} \gsim \omega_{\rm c}^{(+)}$ 
and $\lambda_{-} \lsim \lambda_{+}$ corresponds to 
the system in which the nesting condition of the $+$ band Fermi surface 
is better than that of the $-$ band Fermi surface.

\section{\label{sec:pairinginteraction}
Pairing Interactions 
}

In this section, we examine the transformation of 
the original charge-charge and spin-spin interactions 
into pairing interactions between the electrons on the SFSs. 
We examine interactions of the form 
\Equation{eq:H1chzperp}
{
     H_1 = H_{1 \rm c} + H_{1 z} + H_{1 \perp}, 
     }
with the charge-charge interaction (CI) 
\Equation{eq:H1cdef}
{
     H_{1 \rm c} = \sum_{i,j} V_{ij}^{\rm c} n_i n_j , 
     }
the Ising-type interaction 
\Equation{eq:H1zdef}
{
     H_{1 z} = \sum_{i,j} J_{ij}^{z} S_i^{z} S_j^{z} , 
     }
and the planar spin interaction 
\Equation{eq:H1perpdef}
{
     H_{1 \perp} = \sum_{i,j} J_{ij}^{\perp} 
                      ( S_i^{x} S_j^{x} + S_i^{y} S_j^{y} ) . 
     }
Here, $n_{i} = \sum_{\sigma} c_{i\sigma}^{\dagger} c_{i\sigma}$ and 
$\vS_{i} = \frac{1}{2} \sum_{\sigma_1 \sigma_2} 
c_{i\sigma_1}^{\dagger} {\boldsymbol{\sigma}}_{\sigma_1 \sigma_2} c_{i\sigma_2}$. 
The CI can be derived as an effective interaction mediated by phonons 
and charge fluctuations, 
while the SI can be derived 
as that mediated by spin fluctuations, 
which includes the kinetic exchange and superexchange interactions. 
We have ignored the broken inversion symmetry in the interactions, 
although it might give rise to some interesting effects. 
We define $V^{z} \equiv J^{z}$ and $V^{\perp} \equiv J^{\perp}$ 
for simplicity of the notation. 
The above interactions 
give rise to the pairing interactions 
\Equation{eq:H1pair}
{
     \begin{array}{rcl} 
     H_{1 \rm c} 
     \hsp{-1.0} & = & \hsp{-1.0} \dps{ 
     \frac{1}{N} \hsp{-0.5} \sum_{\vk \vk'\sigma \sigma'} \hsp{-1.0}
       V_{{\rm c}}(\vk,\vk') {\hat \psi}_{\sigma \sigma'}^{\dagger}(\vk) 
                             {\hat \psi}_{\sigma \sigma'}(\vk') 
     }\\[12pt]
     H_{1 z} 
     \hsp{-1.0} & = & \hsp{-1.0} \dps{ 
     \frac{1}{4N} \hsp{-0.5} \sum_{\vk \vk'\sigma \sigma'} \hsp{-1.0}
       \sigma \sigma'
       J_{z}(\vk,\vk') {\hat \psi}_{\sigma \sigma'}^{\dagger}(\vk) 
                       {\hat \psi}_{\sigma \sigma'}(\vk') 
     }\\[12pt]
     H_{1 \perp} 
     \hsp{-1.0} & = & \hsp{-1.0} \dps{ 
       - \frac{1}{2N} \hsp{-0.25} \sum_{\vk \vk'\sigma} \hsp{-0.25}
       J_{\perp}(\vk,\vk') {\hat \psi}_{\sigma -\sigma}^{\dagger}(\vk) 
                       {\hat \psi}_{\sigma -\sigma}(\vk') 
     }
     \end{array}
     }
within the BCS approximation, 
where we have defined 
$$ {
     V^{X}_{\vq}
       \equiv \sum_{ \vR_{ij} } \e^{ - \i \vq \cdot \vR_{ij} } 
         V_{ij}^{X} ~~~ \mbox{($X = c$, $z$, and $\perp$)} , 
     }$$ 
$V_{\rm c}(\vk,\vk') = V_{\vk - \vk'}^{\rm c}$, 
$V_{z}(\vk,\vk') = V_{\vk - \vk'}^{z}$, and 
$V_{\perp}(\vk,\vk') = V_{\vk + \vk'}^{\perp}$. 
The $\alpha$-components $V^{(ss')}_{X \alpha}$ are defined 
by the equations analogous to \eq.{eq:Valpha}. 
Rewriting \eq.{eq:H1pair} into the forms of 
\mbox{\eqs.{eq:H1_in_psi_ss} --
\refeq{eq:VsingVtripdef}}, 
we obtain 
the $\alpha$-components of the singlet and triplet coupling constants 
$$ {
     \begin{array}{rcl} 
     V_{0 \alpha}^{(ss')} 
     & = & \dps{ 
       2 V_{{\rm c}\alpha}^{(ss')} 
       - \frac{1}{2} J_{z \alpha}^{(ss')}
       - J_{\perp \alpha}^{(ss')}
     } \\[8pt] 
     {\tilde V}_{\alpha}^{(ss')} 
     & = & \dps{ 
       \hsp{-1.5} 
       \sum_{\alpha' {\rm (odd)}} \hsp{-0.8} 
       \Bigl [ 
         2 g_{2 \alpha \alpha'}^{(ss')} \, 
           V_{{\rm c} \alpha'}^{(ss')} 
       + \frac{1}{2} 
         {\bar g}_{2 \alpha \alpha'}^{(ss')} 
          J_{z \alpha'}^{(ss')} 
       }\\[8pt] 
     && \dps{ 
        \hsp{6} 
        - g_{2 z \alpha \alpha'}^{(ss')} 
          J_{\perp \alpha'}^{(ss')} 
         \Bigr ]  . 
     }
     \end{array} 
     }$$ 
Therefore, we obtain the transformation rule 
\Equation{eq:GammaTV}
{
     \Gamma_{\alpha}^{(ss')} 
       = 
         \sum_{\alpha ' X} 
           {\cal T}_{X \alpha \alpha'}^{(ss')} V_{X \alpha'}^{(ss')} , 
     }
where 
\Equation{eq:Tdef}
{
     \begin{array}{rcl}
     {\cal T}_{{\rm c} \alpha \alpha'}^{(ss')} 
     \hsp{-0.5} & = & \hsp{-0.5} \dps{ 
       2 s s' \delta_{\alpha \alpha'} 
         + 2 g_{2 \alpha \alpha'}^{(ss')}
     } \\[8pt] 
     {\cal T}_{{z} \alpha \alpha'}^{(ss')} 
     \hsp{-0.5} & = & \hsp{-0.5} \dps{ 
       - \frac{ss'}{2} \delta_{\alpha \alpha'} 
       + \frac{1}{2} {\bar g}_{2 \alpha \alpha'}^{(ss')}
     } \\[8pt] 
     {\cal T}_{{\perp} \alpha \alpha'}^{(ss')} 
     \hsp{-0.5} & = & \hsp{-0.5} \dps{ 
       - s s' \delta_{\alpha \alpha'} 
       - g_{2 z \alpha \alpha'}^{(ss')} , 
     } 
     \end{array}
     }
and 
\Equation{eq:gdef}
{
     \begin{array}{rcl}
     g_{2 \alpha \alpha'}^{(ss')}
     \hsp{-0.5} & = & \hsp{-0.5} \dps{ 
          g_{2 x \alpha \alpha'}^{(ss')} 
        + g_{2 y \alpha \alpha'}^{(ss')} 
        + g_{2 z \alpha \alpha'}^{(ss')} 
         }\\[8pt] 
     {\bar g}_{2 \alpha \alpha'}^{(ss')}
     \hsp{-0.5} & = & \hsp{-0.5} \dps{ 
          g_{2 x \alpha \alpha'}^{(ss')}
        + g_{2 y \alpha \alpha'}^{(ss')}
        - g_{2 z \alpha \alpha'}^{(ss')} 
         }
     \end{array}
     }
with $g_{2 \nu \alpha \alpha'}^{(ss')} 
     = g_{\nu \alpha  \alpha'}^{(s)} 
       g_{\nu \alpha' \alpha }^{(s')}$ 
and 
$$ {
     g_{\nu \alpha \alpha'}^{(s)} 
     = \frac{1}{N} \sum_{\vk} 
       \gamma_{\alpha}^{(s)}(\vk) {\hat g}_{\nu} (\vk) 
       \bigl [ \gamma_{\alpha'}^{(s)}(\vk) \bigr ]^{*} . 
     }$$ 
In particular, for the isotropic spin interaction 
$J_{ij}^{z} = J_{ij}^{\perp} \equiv J_{ij}$, 
we obtain 
\Equation{eq:isotropicspin}
{
     \begin{array}{rcl} 
     V_{0 \alpha}^{(ss')} 
     & = & \dps{ 
       2 V_{{\rm c}\alpha}^{(ss')} 
       - \frac{3}{2} J_{\alpha}^{(ss')}
     } \\[8pt] 
     {\tilde V}_{\alpha}^{(ss')} 
     & = & \dps{ 
       \hsp{-1.5} 
       \sum_{\alpha' {\rm (odd)}} \hsp{-0.8} 
       g_{2 \alpha \alpha'}^{(ss')} \, 
         \Bigl \{ 2 V_{{\rm c} \alpha'}^{(ss')} 
                  + \frac{1}{2} J_{\alpha'}^{(ss')} 
         \Bigr \} . 
     }
     \end{array} 
     }

In the model with a strong on-site Coulomb interaction $U$, the interaction 
\Equation{eq:tJmodelint}
{
     H_1 = \sum_{(i,j)} J \bigl ( \vS_i \cdot \vS_j 
             - \frac{1}{4} n_i n_j \bigr ) 
     }
is derived in the second-order perturbation of 
the hopping integral $t$ with $t \ll U$. 
This form corresponds to 
the present model \eq.{eq:H1chzperp} 
with $J_{ij}^{z} = J_{ij}^{\perp} \equiv J_{ij}$ and 
$V_{ij}^{{\rm c}} = - J_{ij}/4$. 
Therefore, from \eq.{eq:isotropicspin}, 
we obtain 
$$ {
     \begin{array}{rcl} 
     V_{0 \alpha}^{(ss')} 
     & = & \dps{ 
       - 2 J_{\alpha}^{(ss')}
     } \\[4pt] 
     {\tilde V}_{\alpha}^{(ss')} & = & 0 . 
     \end{array} 
     }$$ 
Since \eq.{eq:tJmodelint} does not have triplet interactions, 
no effect due to singlet-triplet mixing, 
which we will describe below, occurs.

In the above equations, $\alpha$ must be of even parity as mentioned above. 
Therefore, 
$\alpha'$ must be of odd parity in the second terms of \eq.{eq:Tdef}, 
because ${\hat \vg}(\vk)$ is an odd function. 
The pairing interactions $V_{X \alpha'}^{(ss')}$ with an odd (even) $\alpha'$ 
contribute to the gap function of even $\alpha$ 
through the triplet (singlet) components 
${\tilde V}_{{\rm tri}}$ ($V_{{\rm sin}}$). 
For example, in a spherically symmetric system, 
a \mbox{p-wave} interaction contributes to 
both \mbox{s-wave} and \mbox{d-wave} pairings, 
while neither \mbox{s-wave} nor \mbox{d-wave} pairings contribute to each other. 

\mbox{}

\mbox{}

\section{\label{sec:limits}
Two Limiting Cases
}

In this section, we compare the results of two opposite limiting cases: 
an equal-band limit and a single-band limit, which are defined below.

\subsection{\label{sec:equalbands}
Equal-band limit 
}

We define the equal-band limit by the conditions for 
the densities of states 
\Equation{eq:rhoequal}
{
     \rho_{+}^{(\alpha)}(0) = \rho_{-}^{(\alpha)}(0) \equiv \rho_{\rm F}, 
     }
and the interaction parameters 
$V_{0\alpha}^{(\pm  \pm)} = V_{0\alpha}^{(\pm \mp)} 
\equiv V_{\alpha}^{\rm sin}$, 
${\tilde V}_{\alpha}^{(\pm \pm)} = {\tilde V}_{\alpha}^{(\pm \mp)} 
\equiv {\tilde V}_{\alpha}^{\rm tri}$, 
and $\omega_{\rm c}^{(+)} = \omega_{\rm c}^{(-)}$, 
which are independent of the band indexes, 
while the SFSs are displaced in momentum space 
because we have set $\langle {\tilde \psi}_{\pm \mp} (\vk) \rangle = 0$. 
We define 
$\lambda_{\alpha}^{\rm sin} 
       = - \frac{1}{2} V_{\alpha}^{\rm sin} \rho_{\rm F}$ 
and 
$\lambda_{\alpha}^{\rm tri} 
       = - \frac{1}{2} {\tilde V}_{\alpha}^{\rm tri} \rho_{\rm F}$. 
Then, we obtain 
${
     \lambda_{\alpha}^{(ss')} 
       = ss' \lambda_{\alpha}^{\rm sin} + \lambda_{\alpha}^{\rm tri} , 
     }$ 
{\it i.e.}, 
$$ {
     \begin{array}{l} 
     \lambda_{\alpha}   \equiv  \lambda_{+} = \lambda_{-} 
       = \lambda_{\alpha}^{\rm sin} + \lambda_{\alpha}^{\rm tri} \\
     \lambda'_{\alpha}  \equiv  \lambda'_{+} = \lambda'_{-} 
       = - \lambda_{\alpha}^{\rm sin} + \lambda_{\alpha}^{\rm tri} . 
     \end{array}
     }$$ 
Setting $\omega_{\rm c} = \omega_{\rm c}^{(\pm)}$ in \eq.{eq:Lambda}, 
we obtain 
\Equation{eq:Lambdaequallimit}
{
     \Lambda_{\alpha} 
       = \lambda_{\alpha} \pm | \lambda'_{\alpha} | 
             = \left \{ 
             \begin{array}{c}
             2 \lambda_{\alpha}^{\rm sin} \\
             2 \lambda_{\alpha}^{\rm tri} 
             \end{array}
             \right . , 
     }
the larger positive one of which is the physical solution. 
For $\Lambda_{\alpha} = 2 \lambda_{\alpha}^{\rm sin}$ 
and $2 \lambda_{\alpha}^{\rm tri}$, 
the gap function becomes 
$$ {
     \left ( 
       \begin{array}{c}
       \Delta_{\alpha}^{(+)} \\ 
       \Delta_{\alpha}^{(-)} 
       \end{array}
     \right ) 
     = 
     \left ( 
       \begin{array}{c}
         \Delta_{\alpha} \\
       - \Delta_{\alpha} 
       \end{array}
     \right ) 
     ~~ \mbox{and} ~~ 
     \left ( 
       \begin{array}{c}
         \Delta_{\alpha} \\
         \Delta_{\alpha} 
       \end{array}
     \right ) , 
     }$$ 
respectively. 
Therefore, in this ideal case, 
the gap function becomes purely singlet or triplet.

\subsection{\label{sec:singleband}
Single-band limit 
}

The single-band limit is defined so that 
only one of the spin-orbit split bands has a Fermi surface, 
which is expressed as 
\Equation{eq:singlelimit}
{
     \rho_{+}^{(\alpha)}(0) = \rho_{\rm F} 
     ~~ \mbox{and} ~~ 
     \rho_{-}^{(\alpha)}(0) = 0 . 
     }
The limit can be used as a theoretical model 
for some of the Fermi surfaces in ${\rm Li_2Pt_3B}$. 
Since $\lambda_{-} = \lambda'_{-} = 0$ and $\Delta_{-} = 0$, 
the linearized gap equation \eq.{eq:lingapeq} becomes 
$\Delta_{+} = \lambda_{+} l_{+} \Delta_{+}$. 
Therefore, we obtain 
\Equation{eq:Lambdasinglelimit}
{
     \Lambda_{\alpha} 
       = \lambda_{\alpha}^{(++)} 
       = \lambda_{\alpha}^{\rm sin} + \lambda_{\alpha}^{\rm tri} , 
     }
with $\omega_{\rm c} = \omega_{\rm c}^{(+)}$. 
The gap function becomes 
$$ {
     \left ( 
       \begin{array}{c}
       \Delta_{\alpha}^{(+)} \\ 
       \Delta_{\alpha}^{(-)} 
       \end{array}
     \right ) 
     = 
     \left ( 
       \begin{array}{c}
         \Delta_{\alpha} \\
         0 
       \end{array}
     \right ) . 
     }$$ 
Hence, we obtain ${\cal F}_{-}(\vk,\omega_n) = 0$ 
and $|\langle {\hat d}(\vk) \rangle| = |\langle {\hat d}_0(\vk) \rangle|$ 
from \eqs.{eq:psilargealpha} and \refeq{eq:GFresults}, 
where $|\cdots|$ is to eliminate arbitrary phase factors. 
In contrast to the equal-band limit, 
the amplitudes of the spin-singlet and triplet components coincide.

Thus, the properties of the superconductivity 
are completely different in the two limits. 
The disappearance of one of the SFSs 
causes drastic changes in the transition temperature 
and gap structure. 
This might explain the difference 
between the experimental results in ${\rm Li_2Pd_3B}$ and ${\rm Li_2Pt_3B}$ 
discussed in \S~1. 
For example, the full-gap structure may change into 
a gap structure with line nodes, 
when the spin-orbit coupling $\alpha$ increases. 
We shall illustrate this in the following sections.

\section{\label{sec:phasediagrams}
Phase Diagrams 
}

We apply the present theory to several specific models. 
As an example, 
we suppose that the system has spherically symmetric Fermi surfaces, 
and ${\hat \vg}(\vk) = {\hat \vk}$. 
We set $\alpha = (l,m)$ and define the spherical harmonic functions as 
$$   
     Y_{lm}({\hat \vk}) = Y_{lm}(\theta_{\hat \vk},\varphi_{\hat \vk}) 
     = P_{l}^{m}(\cos \theta_{\hat \vk}) \e^{\i m \varphi_{\hat \vk}} , 
     $$ 
where $\theta_{\hat \vk}$ and $\varphi_{\hat \vk}$ denote the polar 
angles to express the direction of ${\hat \vk}$, 
and $P_{l}^{m}(w)$ denotes the Legendre polynomial. 
The basis functions are written as 
$$ {
     \gamma_{lm}^{(s)}(\vk) 
       = C_{lm}^{(s)} \, 
         \theta( \omega_{\rm c}^{(s)} - | {\tilde \xi}_{\vk s} | ) \, 
         Y_{lm}({\hat \vk}) 
     }$$ 
with the normalization factor $C_{lm}^{(s)}$. 
In the expansions of the interactions, 
we assume that $V_{X (lm)}^{(ss')} = V_{X l}$ 
and retain the terms up to $l = 2$ for simplicity. 
The coefficients $g_{\nu (lm)(l'm')}^{(ss')}$ 
can be calculated straightforwardly, for example, 
as $  g_{2z(0,0)(1,0)}^{(ss')} 
 = g_{2z(1,0)(0,0)}^{(ss')} = 1/3$, and so on.

In this section, we examine the two limiting cases defined 
in the previous section: 
the equal-band limit and single-band limit. 
The former limit is a simplified model of the system 
in which both SFSs exist [\mbox{case (i)}], 
while the latter limit corresponds to the system 
in which only one of the spin-orbit split bands 
has a Fermi surface [\mbox{case (ii)}].

\subsection{\label{sec:chargephaseD}
Charge-charge interaction 
}

\begin{figure}[hbtp]
\vspace{2ex} 
\vspace{2ex} 
\begin{center}
\includegraphics[width=8.0cm]{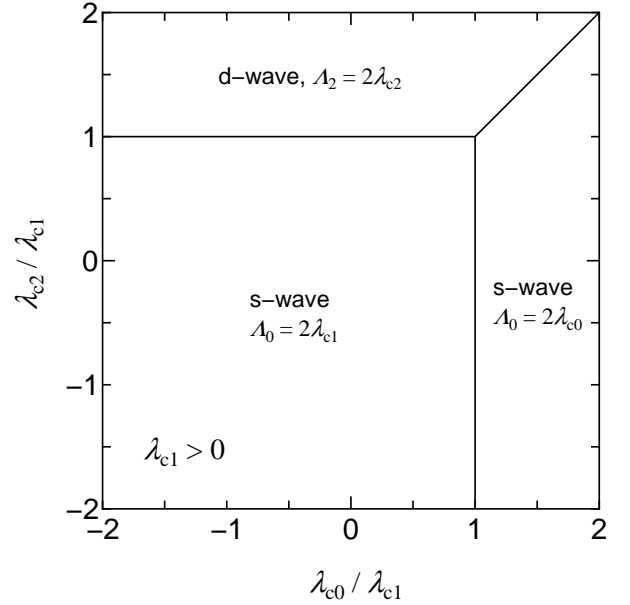}
\end{center}
\caption{Phase diagram of the system with the CI, 
when both of the SFSs exist. 
The spin-triplet interaction is assumed to be attractive. 
}
\label{fig:phasediagram_charge_attrlmd1}
\end{figure}

In the system with the CI defined by \eq.{eq:H1cdef}, 
we obtain 
$$ {
     \begin{array}{rcl}
     \Gamma_{00}^{(ss')} 
     & = & \dps{ 
       2 ss' V_{{\rm c} 0} + 2 V_{{\rm c} 1} 
       } \\
     \Gamma_{2m}^{(ss')} 
     & = & \dps{ 
       2 ss' V_{{\rm c} 2} + \frac{4}{5} V_{{\rm c} 1} 
       } 
     \end{array}
     }$$ 
with $m = \pm 2, \pm 1, 0$. 
Therefore, in the equal-band limit, we obtain 
\Equation{eq:lambda_c}
{
     \begin{array}{rclrcl}
     \lambda_{00}  & = & 
       \dps{   \lambda_{{\rm c}0} + \lambda_{{\rm c}1} , } ~ & ~
     \lambda'_{00} & = & 
       \dps{ - \lambda_{{\rm c}0} + \lambda_{{\rm c}1} , } \\
     \lambda_{2m}  & = & 
       {   \lambda_{{\rm c}2} + \frac{2}{5} \lambda_{{\rm c}1} , } ~ & ~
     \lambda_{2m}' & = & 
       { - \lambda_{{\rm c}2} + \frac{2}{5} \lambda_{{\rm c}1} , }  
     \end{array} 
     }
with $\lambda_{{\rm c}l} = - V_{{\rm c} l} \rho_{\rm F}$. 
Similar results, except the terms including $\lambda_{{\rm c}2}$, 
have been obtained by Samokhin and Mineev~\cite{Sam08}. 
The gap function of the \mbox{s-wave} state has a full-gap structure, 
while those of the \mbox{d-wave} states have line nodes. 
In the present isotropic model, 
the transition temperatures of the \mbox{d-wave} pairing 
are degenerate with respect to $m$, 
and a \mbox{d-wave} state expressed by the linear combination of 
those degenerate states occurs below $\Tc$. 
By minimizing the free energy, 
we obtain a \mbox{d-wave} state with a full-gap structure, 
but this is an artifact due to the isotropy of the model. 
By taking into account the anisotropy that exists in real crystal systems, 
the degeneracy is lifted, 
and some of the solutions with different $m$'s show the highest $\Tc$. 
Therefore, considering the reality, 
we ought to regard the \mbox{d-wave} states as line-node states 
at least near the transition temperature, 
while at low temperatures 
the order parameters with different $m$'s can be mixed 
and the full-gap state may occur. 
From \eqs.{eq:Lambdaequallimit} and \refeq{eq:lambda_c}, 
we obtain 
\Equation{eq:Lambda0andLambda2}
{
     \begin{array}{rcl}
     \Lambda_0 & = & 
       \max ( 2 \lambda_{{\rm c}0}, 2 \lambda_{{\rm 1}}, 0) \\
     \Lambda_2 & = & 
       \max ( 2 \lambda_{{\rm c}2}, 4 \lambda_{{\rm 1}}/5, 0) . 
     \end{array} 
     }
Hence, the resultant coupling constant is expressed as 
\Equation{eq:Lambda_result_c}
{
     \Lambda = \max ( 2 \lambda_{{\rm c}0}, 2 \lambda_{{\rm c}2}, 
                      2 \lambda_{{\rm c}1}, 0 ) , 
     }
which gives the phase diagram shown 
in Fig.~\ref{fig:phasediagram_charge_attrlmd1}. 
The phase diagrams in this paper are not those at $T = 0$, 
but the diagrams of the phases that give $\Tc$. 
Successive transitions to other superconducting phases 
may occur below $\Tc$.

It is found from \eqs.{eq:Lambda0andLambda2} 
and \refeq{eq:Lambda_result_c} that, 
when the even-parity pairing is induced by the spin-triplet pairing, 
it will have the \mbox{s-wave} symmetry rather than the \mbox{d-wave} symmetry, 
because $2 \lambda_{{\rm c}1} > 4 \lambda_{{\rm c}1}/5$ 
for $\lambda_{{\rm c}1} > 0$. 
Although the \mbox{p-wave} attractive interaction contributes to 
both the \mbox{s-wave} pairing and the \mbox{d-wave} pairing, 
the contribution to the \mbox{s-wave} pairing is larger by a factor of $5/2$. 
Consequently, as shown in Fig.~\ref{fig:phasediagram_charge_attrlmd1}, 
when the \mbox{s-wave} and \mbox{d-wave} pairing interactions 
$\lambda_{{\rm c}0}$ and $\lambda_{{\rm c}2}$ are weak or repulsive, 
the \mbox{p-wave} pairing interaction induces 
the \mbox{s-wave} superconductivity. 
Such an \mbox{s-wave} state has a full-gap structure 
similarly to the conventional \mbox{s-wave} state, 
but is at the same time a purely spin-triplet state 
that has the \mbox{d-vector} 
$\langle {\hat \vd}(\vk) \rangle = \langle {\hat d}(\vk) \rangle 
{\hat \vg}(\vk)$ 
with an even parity amplitude $\langle {\hat d}(\vk) \rangle$. 
The gap function becomes 
$\Delta_{\vk s} = - 2 V_{{\rm c} 1} s_{\vk} \langle {\hat d}(\vk) \rangle 
\propto s_{\vk} \theta (\omega_{\rm c}^{(s)} - | \xi_{\vk}^{(s)} | )$, 
which does not have nodes on the Fermi surface 
but has the phase factor $s_{\vk}$. 
The energy gaps of the quasi-particle energies $E_{\vk \pm}$ 
become constants $|\Delta_{(0,0)}^{\pm}|$ independent of ${\hat \vk}$, 
from \eqs.{eq:Eresult} and \refeq{eq:Deltaexpand}. 
The \mbox{s-wave} spin-triplet order is suggested 
in ${\rm Li_2Pt_3B}$ by Yuan {\it et al.}~\cite{Yua06}, 
although ${\rm Li_2Pt_3B}$ is in the opposite limit. 
Interestingly, 
however strong the repulsive spin-singlet interaction is, 
a weaker attractive spin-triplet interaction could cause 
the \mbox{s-wave} superconductivity mentioned above, 
owing to the cancellation effect between the intra- and inter-band 
interactions. 
On the other hand, in Fig.~\ref{fig:phasediagram_charge_attrlmd1}, 
the \mbox{d-wave} phase in the upper area 
and the \mbox{s-wave} phase in the right area 
are conventional pure spin-singlet phases 
induced by \mbox{d-wave} and \mbox{s-wave} pairing interactions, respectively. 
Away from the equal-band limit, 
the mixing of spin-singlet and triplet states occurs.

\begin{figure}[hbtp]
\vspace{2ex} 
\vspace{2ex} 
\begin{center}
\includegraphics[width=8.0cm]{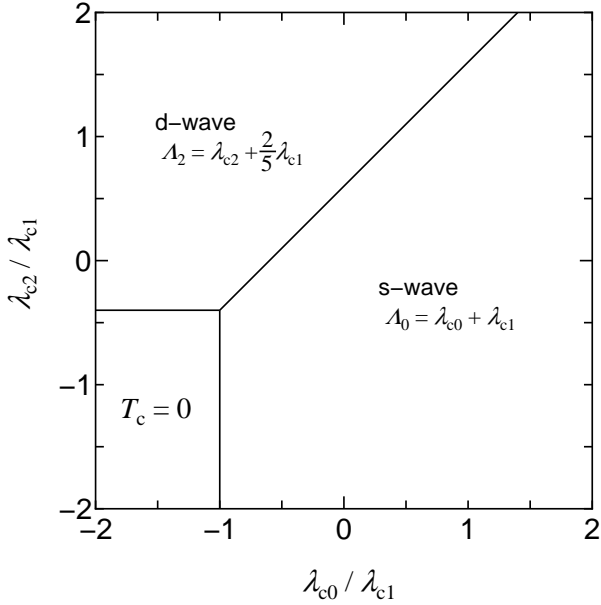}
\end{center}
\caption{Phase diagram of the system with the CI, 
when only one of the spin-orbit split bands has a Fermi surface. 
The spin-triplet interaction is assumed to be attractive. 
}
\label{fig:phasediagram_charge_attrlmd1_single}
\end{figure}

In the single band limit, the effective coupling constant becomes 
\Equation{eq:Lambda_result_c_single}
{
     \Lambda = \max (\Lambda_0,\Lambda_2,0) , 
     }
where $\Lambda_0 = \lambda_{{\rm c}0} + \lambda_{{\rm c}1}$ and 
$\Lambda_2 = \lambda_{2m} 
  = \lambda_{{\rm c}2} + \frac{2}{5} \lambda_{{\rm c}1}$, 
which gives the phase diagram 
shown in Fig.~\ref{fig:phasediagram_charge_attrlmd1_single}. 
In both the \mbox{d-wave} and \mbox{s-wave} phases, 
the singlet-triplet mixing occurs. 
It is found that the \mbox{p-wave} spin-triplet pairing interaction 
stabilizes both the \mbox{d-wave} and \mbox{s-wave} phases, 
but in contrast to the equal band limit, 
the superconductivity is suppressed, 
where both of the \mbox{d-wave} and \mbox{s-wave} interactions are repulsive 
and sufficiently strong.

\begin{figure}[hbtp]
\vspace{2ex} 
\vspace{2ex} 
\begin{center}
\includegraphics[width=8.0cm]{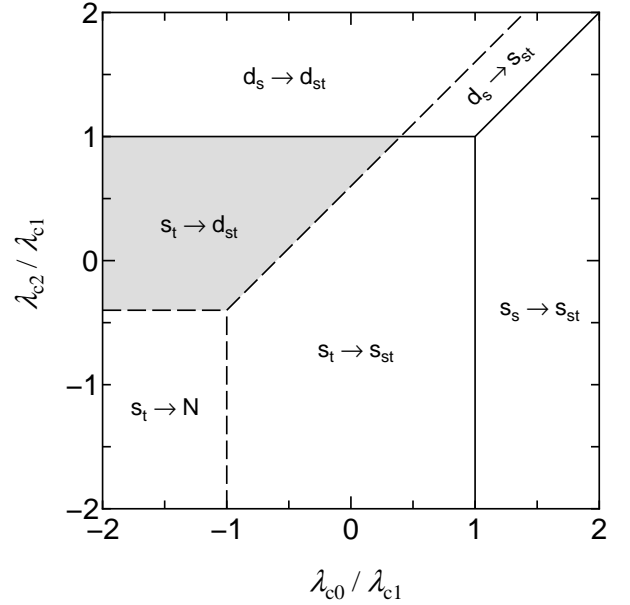}
\end{center}
\caption{Superposition of the phase diagrams 
in Figs.~\ref{fig:phasediagram_charge_attrlmd1} 
and \ref{fig:phasediagram_charge_attrlmd1_single} 
to show the transition when one of the SFSs disappears 
owing to the increase in the spin-orbit coupling. 
The spin-triplet interaction is assumed to be attractive 
($\lambda_{{\rm c}1} > 0$). 
The notation ``${\rm a} \rightarrow {\rm b}$'' means the transition 
from phase ``${\rm a}$'' to phase ``${\rm b}$'', 
where d, s, and N mean the d- and \mbox{s-wave} superconducting phases 
and the normal phase, respectively. 
The suffixes s, t, and st mean the spin-singlet, triplet, and 
singlet-triplet mixed states, respectively. 
For example, ``${\rm s}_{\rm t} \rightarrow {\rm d}_{\rm st}$'' 
means the transition from the \mbox{s-wave} spin-triplet state 
to the \mbox{d-wave} state with both spin-singlet and triplet components. 
}
\label{fig:transition_charge}
\end{figure}

Figure~\ref{fig:transition_charge} shows the transitions 
when one of the SFSs disappears. 
In the gray area, the \mbox{s-wave} full-gap state 
changes into the \mbox{d-wave} state. 
Since the \mbox{d-wave} state can be regarded as 
a line-node state as argued above, 
the transition in this area is a type of FLT. 
This result may explain the observations in 
${\rm Li_2Pd_3B}$ and ${\rm Li_2Pt_3B}$, 
as discussed in \S~\ref{sec:introduction}. 
In this case, the initial state of the transition is 
the \mbox{s-wave} spin-triplet state, 
and the final state is the \mbox{d-wave} state 
with both spin-singlet and triplet components. 
The conventional \mbox{s-wave} spin-singlet state cannot be the initial state 
that changes into the final \mbox{d-wave} state, 
as shown in Fig.~\ref{fig:transition_charge}. 
This result is roughly interpreted as follows. 
For the \mbox{d-wave} state to occur as the final state in the single-band limit, 
the \mbox{s-wave} interaction needs to be weak or repulsive. 
Therefore, if the \mbox{s-wave} state occurs as the initial state 
for the same coupling constants, 
it must be a spin-triplet state induced by the \mbox{p-wave} interaction, 
rather than a spin-singlet state induced by the \mbox{s-wave} interaction. 
This interpretation is not a rigorous proof, 
but is verified by \eqs.{eq:Lambda_result_c} 
and \refeq{eq:Lambda_result_c_single} 
and Fig.~\ref{fig:transition_charge}. 
In most of the gray area, 
the \mbox{s-wave} interaction is repulsive ($\lambda_{{\rm c}0} < 0$), 
while the \mbox{p-wave} and \mbox{d-wave} interactions are attractive; 
the former is stronger than the latter 
($0< \lambda_{{\rm c}2} < \lambda_{{\rm c}1}$). 
These conditions are likely to be satisfied in real materials 
in which both the screened short-range Coulomb repulsion 
and phonon-mediated pairing interactions are strong.

\begin{figure}[hbtp]
\vspace{2ex} 
\vspace{2ex} 
\begin{center}
\includegraphics[width=8.0cm]{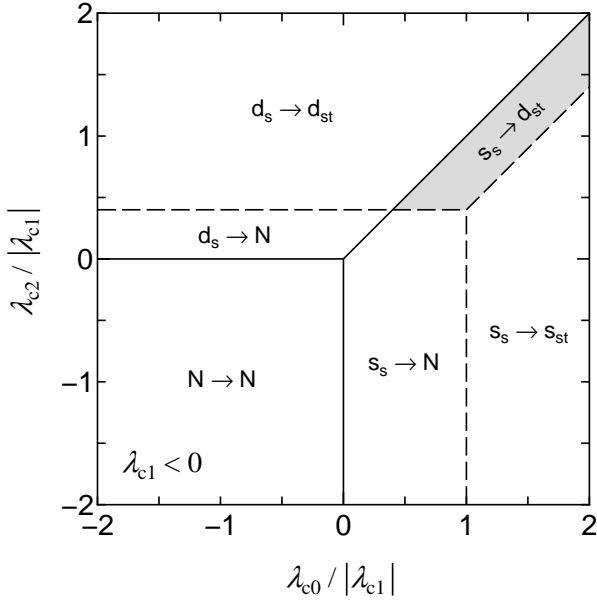}
\end{center}
\caption{Transitions when the interaction is of the charge-charge type 
and the \mbox{p-wave} component is repulsive. 
The legends are the same as those in Fig.~\ref{fig:transition_charge}. 
}
\label{fig:transition_charge_repllmd1}
\end{figure}

Figure~\ref{fig:transition_charge_repllmd1} shows the transitions 
in the case that the \mbox{p-wave} component of the interaction is repulsive, 
which might be unlikely if we suppose a phonon-mediated pairing interaction 
as the CI~\cite{Fou77}. 
In the gray area, the \mbox{s-wave} full-gap state changes into the \mbox{d-wave} state. 
In contrast to the previous case, 
the initial \mbox{s-wave} state is of spin-singlet pairing.

\subsection{\label{sec:spinphaseD}
Spin-spin interactions 
}

Next, we examine the phase diagrams and the transitions 
of the systems with the SIs $H_{z 1}$ and $H_{\perp 1}$ 
of \eqs.{eq:H1zdef} and \refeq{eq:H1perpdef}. 
In this type of interaction, anisotropies of interactions 
in the spin space play an essential role. 
Depending on the anisotropy and the sign of the interactions, 
the transition from the full-gap state to the line-node state (FLT) 
can occur.

\subsubsection{\label{sec:IsingspinphaseD}
Ising-type interaction 
}

In the system with only the Ising-type interaction \eq.{eq:H1zdef}, 
we obtain 
$$ {
     \begin{array}{rcl}
     \Gamma_{00}^{(ss')} 
     & = & { 
       - \frac{1}{2} ss' J_{z 0} + \frac{1}{6} J_{z 1} 
       } \\[4pt]
     \Gamma_{20}^{(ss')} 
     & = & { 
       - \frac{1}{2} ss' J_{z 2} - \frac{1}{15} J_{z 1} 
       } \\[4pt]
     \Gamma_{2,\pm 1}^{(ss')} 
     & = & { 
       - \frac{1}{2} ss' J_{z 2} 
       } \\[4pt]
     \Gamma_{2,\pm 2}^{(ss')} 
     & = & { 
       - \frac{1}{2} ss' J_{z 2} + \frac{1}{5} J_{z 1} 
       } . 
     \end{array}
     }$$ 
(i) In the equal-band limit, we obtain 
\Equation{eq:lambda_z}
{
     \begin{array}{rclrcl}
     \lambda_{00}
     & = & { 
         \lambda_{z0} - \frac{1}{3} \lambda_{z1} , } 
       ~ & ~
     \lambda_{00}'
     & = & { 
       - \lambda_{z0} - \frac{1}{3} \lambda_{z1} , 
           } \\[4pt]
     \lambda_{20}
     & = & { 
         \lambda_{z2} + \frac{2}{15} \lambda_{z1} , }
       ~ & ~
     \lambda_{20}'
     & = & { 
       - \lambda_{z2} + \frac{2}{15} \lambda_{z1} , 
           } \\[4pt]
     \lambda_{2 \pm 1}
     & = & {   \lambda_{z2} , }
       ~ & ~
     \lambda_{2 \pm 1}'
     & = & { - \lambda_{z2} , 
           } \\[4pt]
     \lambda_{2 \pm 2}
     & = & { 
         \lambda_{z2} - \frac{4}{5} \lambda_{z1} , }
       ~ & ~
     \lambda_{2 \pm 2}'
     & = & { 
       - \lambda_{z2} - \frac{4}{5} \lambda_{z1} , } \\
     \end{array}
     }
with $\lambda_{z l} = \frac{1}{4} J_{z l} \rho_{\rm F}$. 
Hence, we obtain 
\Equation{eq:Lambda_z_lm}
{
     \begin{array}{rcl} 
     \Lambda_{0} 
       & = & 
         \max( 2 \lambda_{z0} , - \frac{2}{3} \lambda_{z1} , 0 ) , \\[4pt] 
     \Lambda_{20} 
       & = & 
         \max( 2 \lambda_{z2} , \frac{4}{15} \lambda_{z1} , 0 ) , \\[4pt] 
     \Lambda_{2,\pm 1} 
       & = & 
         \max( 2 \lambda_{z2} , 0 ) , \\[4pt] 
     \Lambda_{2,\pm 2} 
       & = & 
         \max( 2 \lambda_{z2} , - \frac{8}{5} \lambda_{z1} , 0 ) , 
     \end{array}
     } 
and 
\Equation{eq:Lambda_z}
{
     \Lambda = \max ( 2 \lambda_{z0}, 
                      2 \lambda_{z2}, 
                      - \frac{8}{ 5} \lambda_{z1}, 
                        \frac{4}{15} \lambda_{z1}, 0) . 
     }
Interestingly, 
the {\it repulsive} \mbox{p-wave} spin-triplet interaction stabilizes 
both the \mbox{d-wave} spin-triplet state with $(l,m) = (2,\pm 2)$ 
and the \mbox{s-wave} spin-triplet state. 
However, because of the numerical factors in front of $\lambda_{z1}$, 
the former overcomes the latter. 
(ii) In the single-band limit, we obtain 
\Equation{eq:Lambda_z_lm_single}
{
     \begin{array}{rcl}
     \Lambda_{0}
     & = & { 
         \lambda_{z0} - \frac{1}{3} \lambda_{z1} , 
           } \\[4pt]
     \Lambda_{20}
     & = & { 
         \lambda_{z2} + \frac{2}{15} \lambda_{z1} , 
           } \\[4pt]
     \Lambda_{2,\pm 1}
     & = & {   \lambda_{z2} , 
           } \\[4pt]
     \Lambda_{2,\pm 2}
     & = & { 
         \lambda_{z2} - \frac{4}{5} \lambda_{z1} } . 
     \end{array}
     }
The resultant coupling constant is 
\Equation{eq:Lambda_z_single}
{
     \Lambda 
       = \max ( \Lambda_0, \Lambda_{20}, 
                \Lambda_{2,\pm 1}, \Lambda_{2, \pm 2},0) . 
     }

\subsubsection{\label{sec:spinxyphaseD}
Planar spin interaction 
}

In the system with only the planar spin interaction \eq.{eq:H1perpdef}, 
we obtain 
$$ {
     \begin{array}{rcl}
     \Gamma_{00}^{(ss')} 
     & = & { 
       - ss' J_{\perp 0} + \frac{1}{3} J_{\perp 1} , 
       } \\[4pt]
     \Gamma_{20}^{(ss')} 
     & = & { 
       - ss' J_{\perp 2} + \frac{4}{15} J_{\perp 1} , 
       } \\[4pt]
     \Gamma_{2,\pm 1}^{(ss')} 
     & = & { 
       - ss' J_{\perp 2} + \frac{1}{5} J_{\perp 1} , 
       } \\[4pt]
     \Gamma_{2,\pm 2}^{(ss')} 
     & = & { 
       - ss' J_{\perp 2} 
       } . 
     \end{array}
     }$$ 
(i) In the equal-band limit, we obtain 
$$ {
     \begin{array}{rclrcl}
     \lambda_{00}
     & = & { 
         \lambda_{\perp 0} - \frac{1}{3} \lambda_{\perp 1} , 
           }
       ~ & ~
     \lambda_{00}'
     & = & { 
       - \lambda_{\perp 0} - \frac{1}{3} \lambda_{\perp 1} , 
           } \\[4pt]
     \lambda_{20}
     & = & { 
         \lambda_{\perp 2} - \frac{4}{15} \lambda_{\perp 1} , 
           }
       ~ & ~
     \lambda_{20}'
     & = & { 
       - \lambda_{\perp 2} - \frac{4}{15} \lambda_{\perp 1} , 
           } \\[4pt]
     \lambda_{2 \pm 1}
     & = & {   
         \lambda_{\perp 2} - \frac{1}{5} \lambda_{\perp 1} , 
           }
       ~ & ~
     \lambda_{2 \pm 1}'
     & = & { 
       - \lambda_{\perp 2} - \frac{1}{5} \lambda_{\perp 1} , 
           } \\[4pt]
     \lambda_{2 \pm 2}
     & = & { 
         \lambda_{\perp 2} , 
           } 
       ~ & ~
     \lambda_{2 \pm 2}'
     & = & { 
       - \lambda_{\perp 2} , 
           } 
     \end{array}
     }$$ 
with $\lambda_{\perp l} = \frac{1}{2} J_{\perp l} \rho_{\rm F}$. 
Hence, we obtain 
$$ {
     \begin{array}{rcl} 
     \Lambda_{0} 
       & = & 
         \max( 2 \lambda_{\perp 0} , - \frac{2}{3} \lambda_{\perp 1} , 0 ) , 
         \\[4pt] 
     \Lambda_{20} 
       & = & 
         \max( 2 \lambda_{\perp 2} , - \frac{8}{15} \lambda_{\perp 1} , 0 ) , 
         \\[4pt] 
     \Lambda_{2,\pm 1} 
       & = & 
         \max( 2 \lambda_{\perp 2} , - \frac{2}{5} \lambda_{\perp 1}, 0 ) , 
         \\[4pt] 
     \Lambda_{2,\pm 2} 
       & = & 
         \max( 2 \lambda_{\perp 2} , 0 ) , 
     \end{array}
     }$$ 
and 
$$ {
     \Lambda = \max ( 2 \lambda_{\perp 0}, 
                      2 \lambda_{\perp 2}, 
                      - \frac{2}{3} \lambda_{\perp 1}, 0) . 
     }$$ 
(ii) In the single-band limit, we obtain 
$$ {
     \begin{array}{rcl}
     \Lambda_{0}
     & = & { 
         \lambda_{\perp 0} - \frac{1}{3} \lambda_{\perp 1} , 
           } \\[4pt]
     \Lambda_{20}
     & = & { 
         \lambda_{\perp 2} - \frac{4}{15} \lambda_{\perp 1} , 
           } \\[4pt]
     \Lambda_{2,\pm 1}
     & = & {   \lambda_{\perp 2} - \frac{1}{5} \lambda_{\perp 1} , 
           } \\[4pt]
     \Lambda_{2,\pm 2}
     & = & { 
         \lambda_{\perp 2} } . 
     \end{array}
     }$$ 
The resultant coupling constant is the maximum positive one, 
as given by \eq.{eq:Lambda_z_single}.

\subsubsection{\label{sec:isotropicspinphaseD}
Isotropic spin interaction 
}

Lastly, in the system with the isotropic spin interaction 
$J_{ij}^{z} = J_{ij}^{\perp} \equiv J_{ij}$, 
we obtain 
$$ {
     \begin{array}{rcl}
     \Gamma_{0}^{(ss')} 
     & = & { 
       - ss' \frac{3}{2} J_{0} + \frac{1}{2} J_{1} , 
       } \\[4pt]
     \Gamma_{2}^{(ss')} 
     & = & { 
       - ss' \frac{3}{2} J_{2} + \frac{1}{5} J_{1} . 
       } 
     \end{array}
     }$$ 
(i) In the equal-band limit, we obtain 
$$ {
     \begin{array}{rclrcl}
     \lambda_{0}
     & = & { 
         \lambda_{0} - \frac{1}{3} \lambda_{1} , 
           }
       ~ & ~
     \lambda_{0}'
     & = & { 
       - \lambda_{0} - \frac{1}{3} \lambda_{1} , 
           } \\[4pt]
     \lambda_{2}
     & = & { 
         \lambda_{2} - \frac{2}{15} \lambda_{1} , 
           }
       ~ & ~
     \lambda_{2}'
     & = & { 
       - \lambda_{2} - \frac{2}{15} \lambda_{1} . 
           } 
     \end{array}
     }$$ 
with $\lambda_{l} = \frac{3}{4} J_{l} \rho_{\rm F}$. 
Hence, we obtain 
$$ {
     \begin{array}{rcl} 
     \Lambda_{0} 
       & = & 
         \max( 2 \lambda_{0} , - \frac{2}{3} \lambda_{1} , 0 ) , 
         \\[4pt] 
     \Lambda_{2} 
       & = & 
         \max( 2 \lambda_{2} , - \frac{4}{15} \lambda_{1} , 0 ) , 
     \end{array}
     }$$ 
and 
\Equation{eq:Lambda_isoJ}
{
     \Lambda = \max ( 2 \lambda_{0}, 
                      2 \lambda_{2}, 
                      - \frac{2}{3} \lambda_{1}, 0) . 
     }
(ii) In the single-band limit, 
we obtain 
$$ {
     \begin{array}{rcl}
     \Lambda_{0}
     & = & { 
         \lambda_{0} - \frac{1}{3} \lambda_{1} , 
           } \\[4pt]
     \Lambda_{2}
     & = & { 
         \lambda_{2} - \frac{2}{15} \lambda_{1} , 
           } 
     \end{array}
     }$$ 
and 
\Equation{eq:Lambda_isoJ_single}
{
     \Lambda = \max ( 
         \lambda_{0} - \frac{1}{3} \lambda_{1} , 
         \lambda_{2} - \frac{2}{15} \lambda_{1} , 0 ) . 
     }

\subsubsection{\label{sec:spinPhaseD}
Phase diagrams and transitions 
}

\begin{figure}[hbtp]
\vspace{2ex} 
\vspace{2ex} 
\begin{center}
\includegraphics[width=8.0cm]{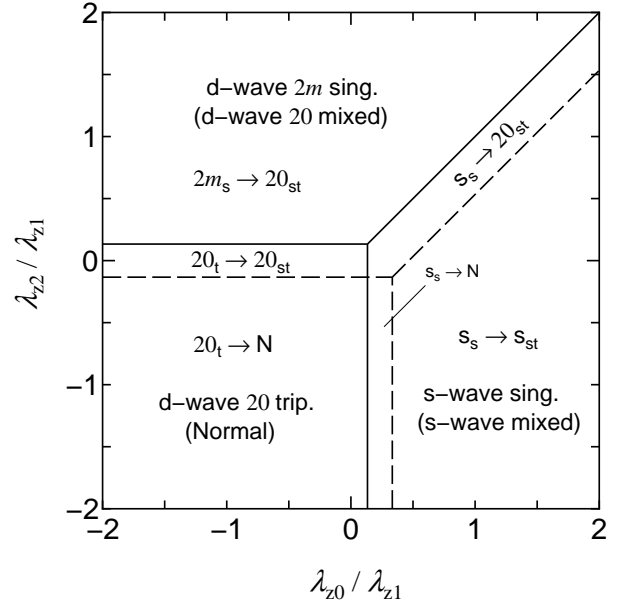}
\end{center}
\caption{Phase diagrams and transitions 
for Ising-type interaction with attractive \mbox{p-wave} components. 
Solid and broken lines are the phase boundaries 
in cases (i) and (ii), respectively. 
The states in the latter case are shown in the brackets. 
The notation ``2$m$'' means the \mbox{d-wave} state 
with the quantum number $(2,m)$ (see the text). 
The other legends are as shown 
in the caption of Fig.~\ref{fig:transition_charge}. 
}
\label{fig:transition_Ising_attrlmd1}
\end{figure}

\begin{figure}[hbtp]
\vspace{2ex} 
\vspace{2ex} 
\begin{center}
\includegraphics[width=8.0cm]{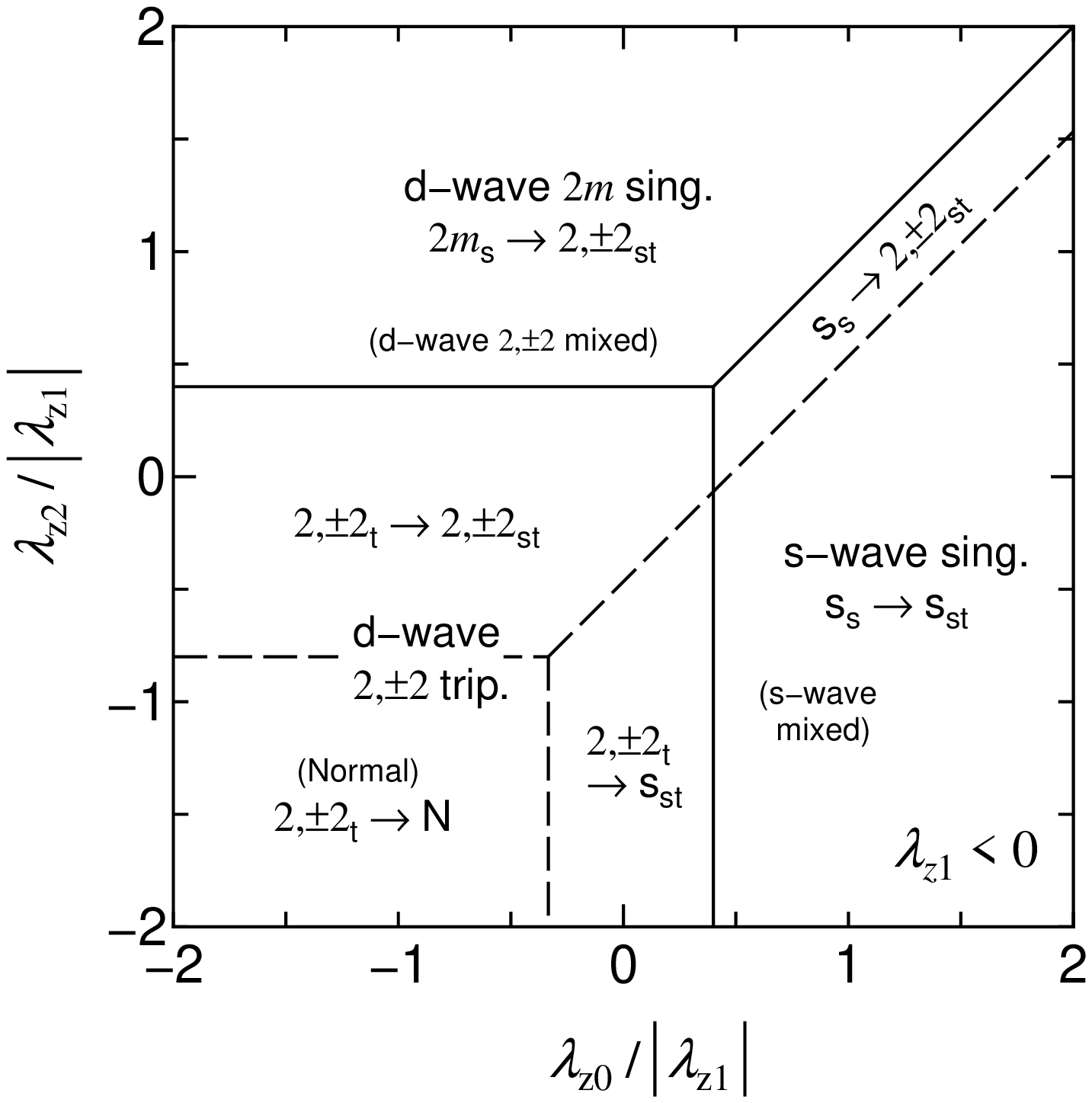}
\end{center}
\caption{Phase diagrams and transitions 
for Ising-type interaction with repulsive \mbox{p-wave} components. 
The legends are as shown in the captions of 
Figs.~\ref{fig:transition_charge}~and~\ref{fig:transition_Ising_attrlmd1}. 
}
\label{fig:transition_Ising_repllmd1}
\end{figure}

\begin{figure}[hbtp]
\vspace{2ex} 
\vspace{2ex} 
\begin{center}
\includegraphics[width=8.0cm]{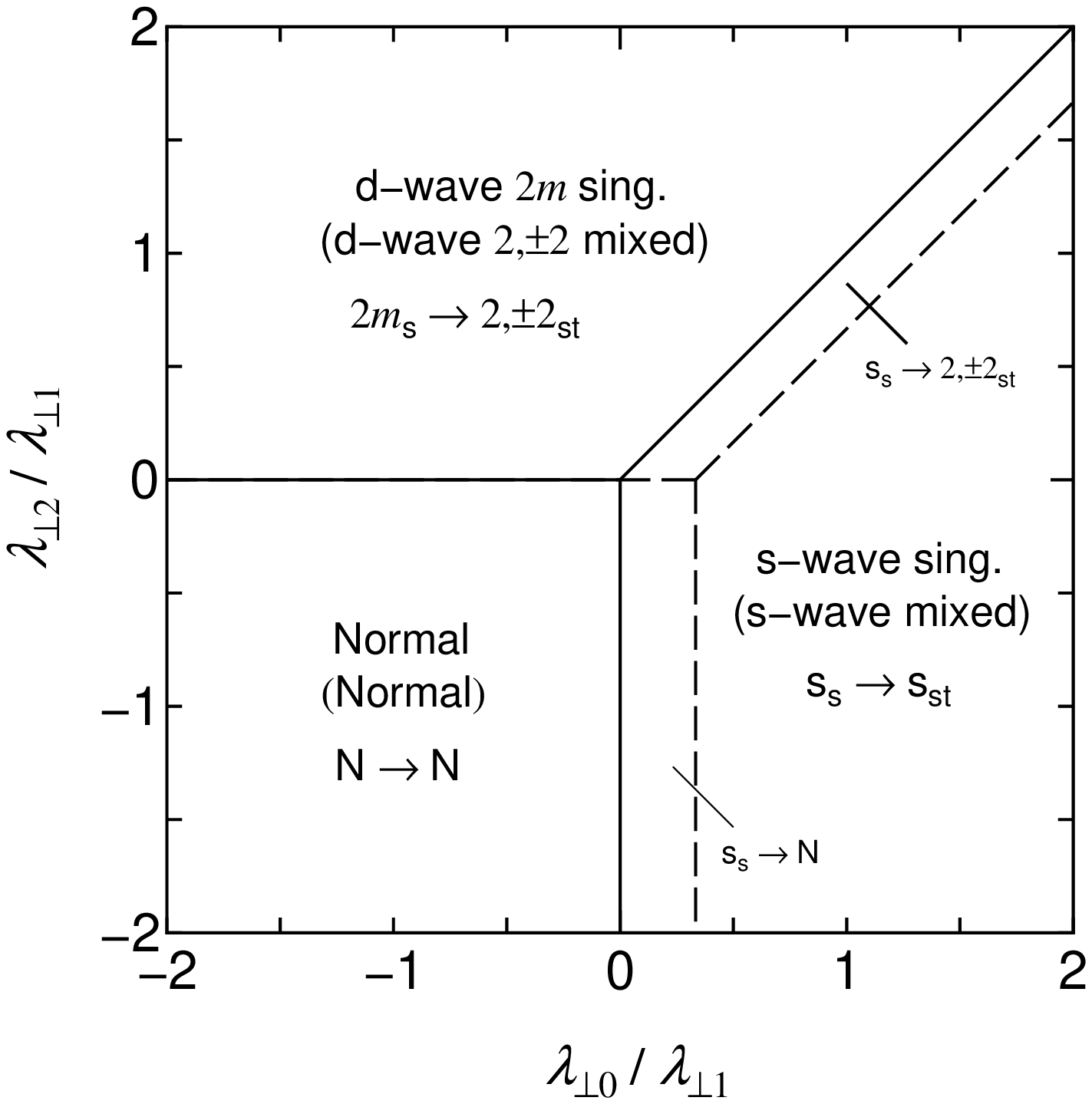}
\end{center}
\caption{Phase diagrams and transitions 
for planar spin interaction with attractive \mbox{p-wave} components. 
The legends are as shown in the captions of 
Figs.~\ref{fig:transition_charge}~and~\ref{fig:transition_Ising_attrlmd1}. 
}
\label{fig:transition_planar_attrlmd1}
\end{figure}

\begin{figure}[hbtp]
\vspace{2ex} 
\vspace{2ex} 
\begin{center}
\includegraphics[width=8.0cm]{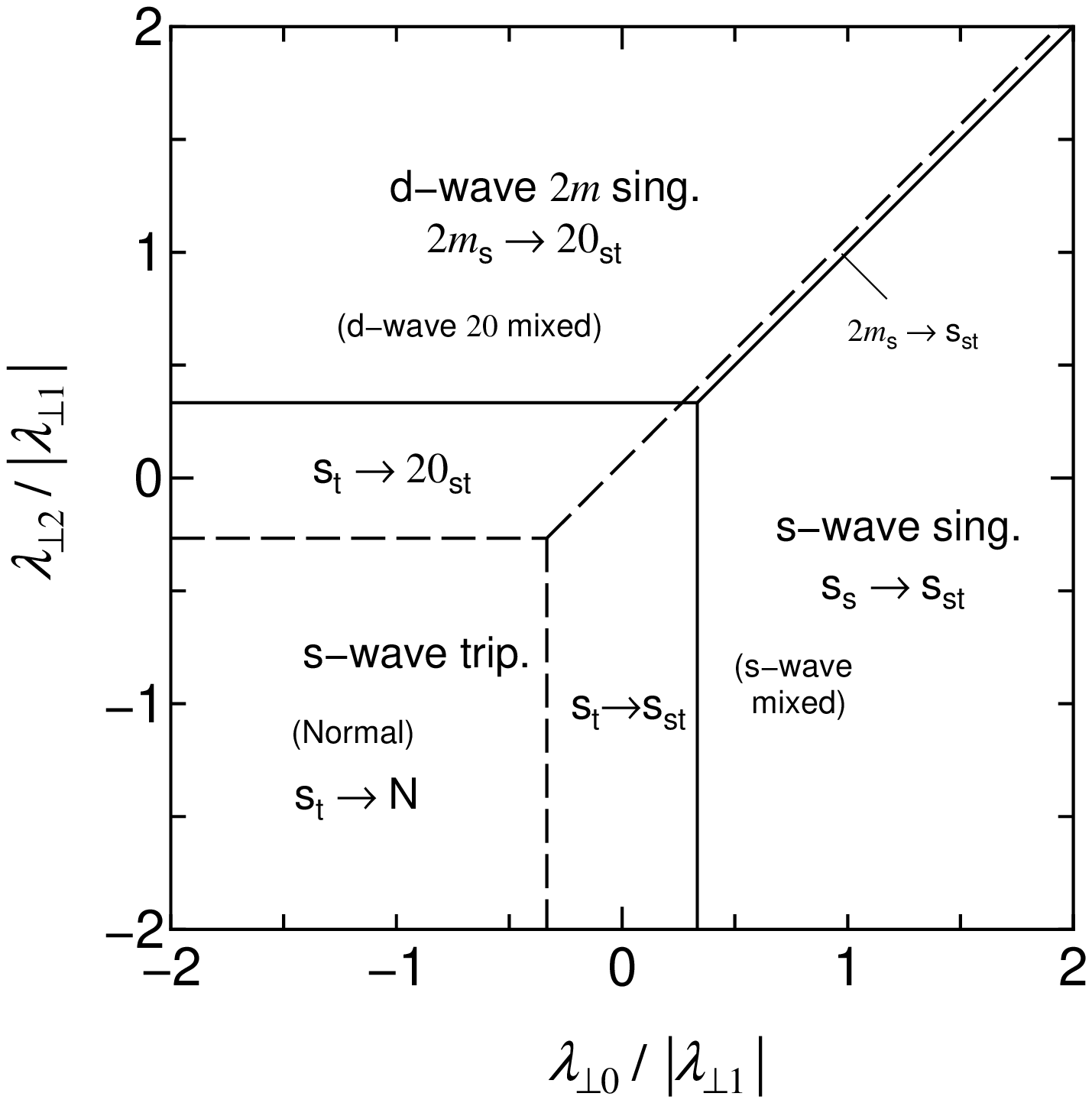}
\end{center}
\caption{Phase diagrams and transitions 
for planar spin interaction with repulsive \mbox{p-wave} components. 
The legends are as shown in the captions of 
Figs.~\ref{fig:transition_charge}~and~\ref{fig:transition_Ising_attrlmd1}. 
}
\label{fig:transition_planar_repllmd1}
\end{figure}

\begin{figure}[hbtp]
\vspace{2ex} 
\vspace{2ex} 
\begin{center}
\includegraphics[width=8.0cm]{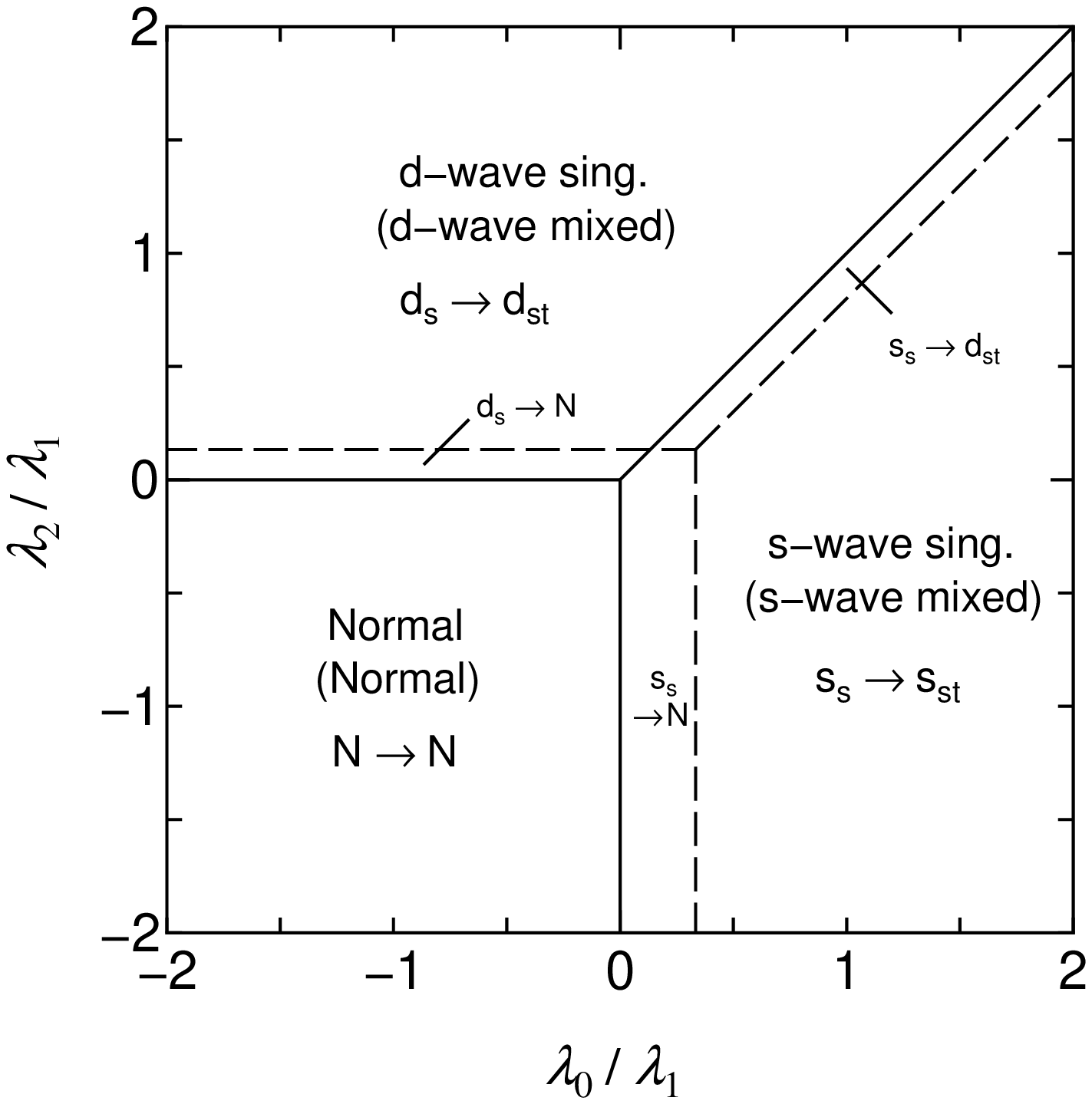}
\end{center}
\caption{Phase diagrams and transitions 
for isotropic spin interaction with attractive \mbox{p-wave} components. 
The legends are as shown in the captions of 
Figs.~\ref{fig:transition_charge}~and~\ref{fig:transition_Ising_attrlmd1}. 
}
\label{fig:transition_iso_attrlmd1}
\end{figure}

\begin{figure}[hbtp]
\vspace{2ex} 
\vspace{2ex} 
\begin{center}
\includegraphics[width=8.0cm]{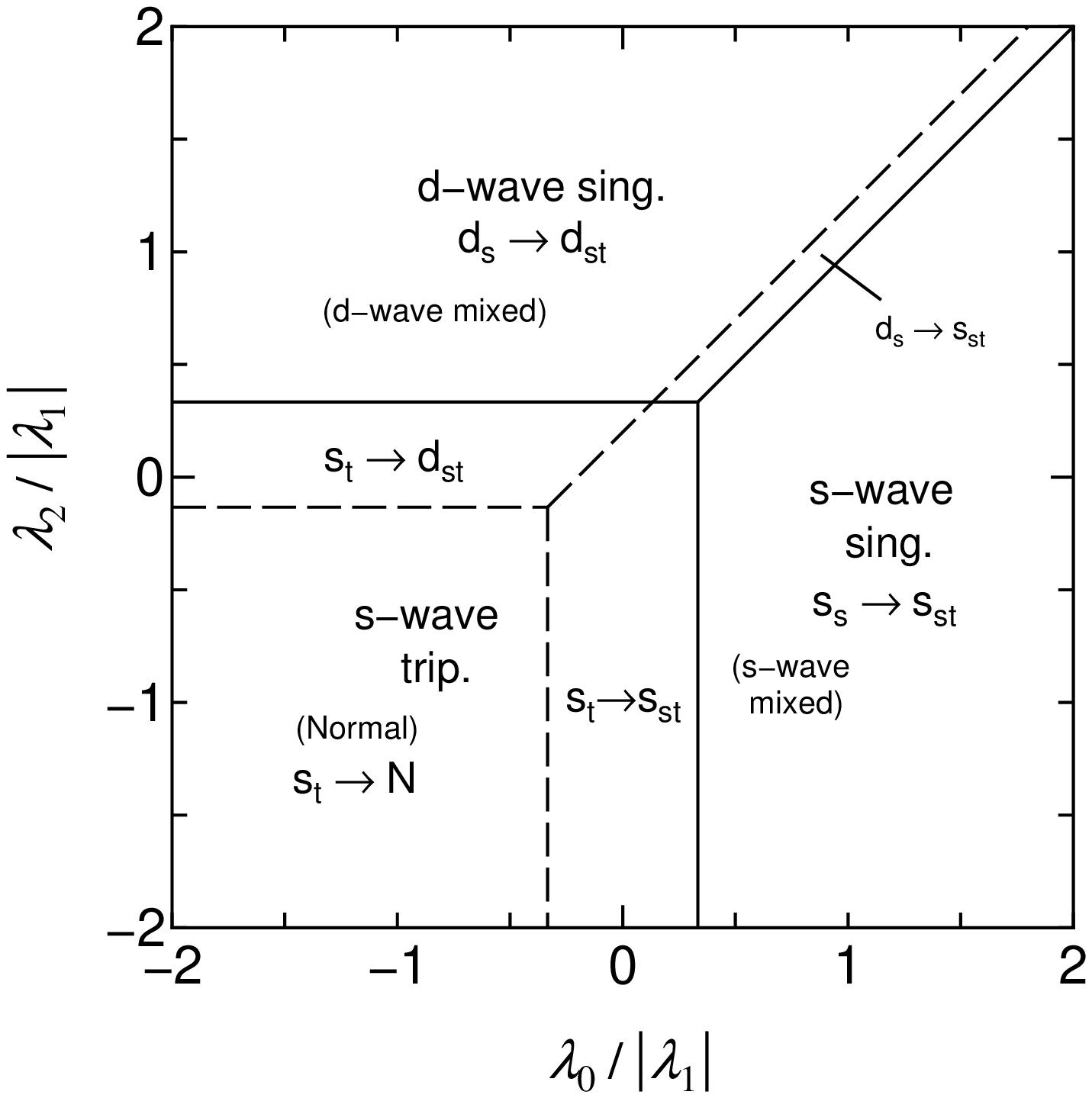}
\end{center}
\caption{Phase diagrams and transitions 
for isotropic spin interaction with repulsive \mbox{p-wave} components. 
The legends are as shown in the captions of 
Figs.~\ref{fig:transition_charge}~and~\ref{fig:transition_Ising_attrlmd1}. 
}
\label{fig:transition_iso_repllmd1}
\end{figure}

In this subsection, 
we examine the phase diagrams of the systems with SIs. 
Figures~\ref{fig:transition_Ising_attrlmd1}
      - \ref{fig:transition_iso_repllmd1} 
show the phase diagrams of the systems with various types of SIs 
(see Table~\ref{table:spinphased}). 
In each figure, two phase diagrams are superposed as shown in 
Figs.~\ref{fig:phasediagram_charge_attrlmd1}~-~\ref{fig:transition_charge} 
for the CI. 
The solid lines show the phase boundaries in \mbox{case (i)}, 
and the broken lines and the texts in brackets 
show the phase boundaries 
and the symmetries of the phases in \mbox{case (ii)}, respectively. 
As explained in the caption of Fig.~\ref{fig:transition_charge}, 
the notation ``${\rm a} \rightarrow {\rm b}$'' means the transition 
from phase ``a'' to phase ``b'', 
when one of the SFSs disappears [case (i) $\rightarrow$ case (ii)]. 
In addition to the characters ``s'', ``d'', and ``N'', 
we have defined the notation ``2$m$'' that means 
the \mbox{d-wave} state with $\Delta \propto Y_{2m}$.

In case (i), anomalous even-parity (s- and \mbox{d-wave}) 
spin-triplet superconducting phases occur, 
where both the s- and \mbox{d-wave} components of the interactions are repulsive. 
Ising-type interactions induce \mbox{d-wave} spin-triplet states 
with different $m$'s for either sign of the \mbox{p-wave} component 
(Figs.~\ref{fig:transition_Ising_attrlmd1} and 
      ~\ref{fig:transition_Ising_repllmd1}), 
in contrast to the CI. 
The magnetically mediated pairing interactions 
could also induce \mbox{s-wave} spin-triplet states, 
when the interaction is planar or isotropic and 
their \mbox{p-wave} component is repulsive 
(Figs.~\ref{fig:transition_planar_attrlmd1}
- \ref{fig:transition_iso_repllmd1}). 
These results are quite different from that for the system with the CI, 
in which an \mbox{s-wave} spin-triplet state occurs 
only when the \mbox{p-wave} component of the interaction is attractive. 
These differences are explained as follows. 
When singlet interactions are repulsive, 
only a triplet \mbox{p-wave} interaction contributes to 
the even-parity state 
through the mixing effect due to spin-orbit coupling. 
Depending on the sign of the second terms of the matrix elements 
${\cal T}_{X \alpha \alpha'}^{(ss')}$ ($X = {\rm c}, z, \perp$) 
defined by \eq.{eq:Tdef}, 
an attractive or repulsive \mbox{p-wave} interaction 
may contribute to the superconductivity of the anomalous type. 
Independently of the type of the interactions, 
such an anomalous phase disappears, 
when one of the SFSs disappears owing to the stronger spin-orbit coupling.

\begin{table}[hbtp]
\vspace{2ex} 
\vspace{2ex} 
\begin{center}
\caption{
Table of the phase diagrams. 
}
\label{table:spinphased}
\begin{tabular}{c}
\\[-10pt]
\begin{tabular}{l|c|p{8ex}|p{8ex}}
\raisebox{-0.0ex}{Type of} 
  & \raisebox{-0.0ex}{coupling} 
                & \multicolumn{2}{c}{p-wave component } \\ 
\cline{3-4} 
\raisebox{0.0ex}{interaction} 
  & \raisebox{0.0ex}{constant} 
  & $\lambda_1 > 0 $  &  $\lambda_1 < 0$ \\ 
\hline 
Ising           & $J_z \ne 0$, $J_{\perp} = 0$ 
                & Fig.~\ref{fig:transition_Ising_attrlmd1} 
                & Fig.~\ref{fig:transition_Ising_repllmd1} \\ 
Planar          & $J_z = 0$, $J_{\perp} \ne 0$ 
                & Fig.~\ref{fig:transition_planar_attrlmd1} 
                & Fig.~\ref{fig:transition_planar_repllmd1} \\ 
Isotropic       & $J_z = J_{\perp} \equiv J $ 
                & Fig.~\ref{fig:transition_iso_attrlmd1} 
                & Fig.~\ref{fig:transition_iso_repllmd1} \\ 
\end{tabular}
\end{tabular}
\end{center}
\end{table}

\subsubsection{\label{sec:fullgaplinenodetr}
Transitions from the full-gap state 
to the line-node state 
}

The FLT due to the disappearance of one of the SFSs also occurs for the SI, 
as shown in \mbox{Figs.~\ref{fig:transition_Ising_attrlmd1} 
- \ref{fig:transition_iso_repllmd1}}, 
which are summarized in Table~\ref{table:FLT}. 
Practically, we can regard \mbox{d-wave} states as line-node states, 
even when they degenerate with respect to $m$, 
by the argument in \S~\ref{sec:chargephaseD}.

\begin{table}[hbtp]
\vspace{2ex} 
\vspace{2ex} 
\begin{center}
\caption{
Relation between the type of interaction 
and the signs of the coupling constants for the FLT to occur. 
In each phase diagram, 
the signs are those for the major part of the area in which the FLT occurs. 
The double sign $\pm$ means that $\lambda_2$ can take 
either sign, but the absolute value is small. 
``${\rm s_s}$'' and ``${\rm s_t}$'' denote the \mbox{s-wave} 
spin-singlet and triplet states, respectively. 
}
\label{table:FLT}
\begin{tabular}{c}
\\[-10pt]
\hsp{-2.0} 
\begin{tabular}{l|c|c|c|c|l}
\raisebox{-0.0ex}{Type of} 
  & \multicolumn{3}{c|}{Sign of $\lambda_l$ } 
  & \raisebox{-0.0ex}{FLT} 
  & \raisebox{-0.0ex}{Phase} \\
\cline{2-4}
\raisebox{0.0ex}{interaction}
  & ~s~   & ~p~   & ~d~  
  & ${\rm s} \rightarrow {\rm d} \, (l,m)$  
  & \raisebox{0.0ex}{diagram} \\ 
\hline 
Charge                    & $-$ & $+$ & $+$ 
  & ${\rm s_t} \rightarrow {\rm d} \, (2,m)$ 
  & Fig.~\ref{fig:transition_charge} \\
                          & $+$ & $-$ & $+$ 
  & ${\rm s_s} \rightarrow {\rm d} \, (2,m) $ 
  & Fig.~\ref{fig:transition_charge_repllmd1} \\
Ising spin                & $+$ & $+$ & $+$ 
  & ${\rm s_s} \rightarrow {\rm d} \, (2,0)$ 
  & Fig.~\ref{fig:transition_Ising_attrlmd1} \\
                          & $+$ & $-$ & $+$ 
  & \!\! 
    ${\rm s_s} \rightarrow {\rm d} \, (2,\pm 2 ) $ \!\! 
  & Fig.~\ref{fig:transition_Ising_repllmd1} \\
Planar spin               & $+$ & $+$ & $+$ 
  & \!\! 
    ${\rm s_s} \rightarrow {\rm d} \, (2,\pm 2)$ \!\! 
  & Fig.~\ref{fig:transition_planar_attrlmd1} \\ 
                          & $-$ & $-$ & $\pm$ 
  & ${\rm s_t} \rightarrow {\rm d} \, ( 2,0 ) $ 
  & Fig.~\ref{fig:transition_planar_repllmd1} \\ 
Isotropic spin            & $+$ & $+$ & $+$ 
  & ${\rm s_s} \rightarrow {\rm d} \, (2,m) $ 
  & Fig.~\ref{fig:transition_iso_attrlmd1} \\ 
                          & $-$ & $-$ & $\pm$ 
  & ${\rm s_t} \rightarrow {\rm d} \, (2,m)$ 
  & Fig.~\ref{fig:transition_iso_repllmd1} \\ 
\end{tabular}
\end{tabular}
\end{center}
\end{table}

In these phase diagrams, the regions where the FLT occurs are narrower 
than that in the phase diagram (Fig.~\ref{fig:transition_charge}) for the CI. 
Furthermore, it seems difficult that the \mbox{s-wave} component 
becomes attractive for interactions of magnetic origin. 
If we exclude such a situation, 
the remaining possibilities are planar and 
isotropic spin interactions with $\lambda_1 < 0$. 
However, in such cases, the FLT occurs only in small regions, 
in which the pairing interaction is very weak. 
Therefore, in the present theory, if the FLT occurs, 
it is most likely that the CI is the most dominant pairing interaction.

\section{\label{sec:last}
Summary and Discussion 
}

We have examined the superconductivity 
in noncentrosymmetric systems 
with various types of interactions between electrons. 
We have presented a formulation of the superconductivity, 
and obtained the transition temperatures and gap functions, 
including the results that have been obtained 
by previous authors~\cite{Gor01,Ser04,Fri04,Sam08}. 
We have derived the pairing interaction \eq.{eq:GammaTV} 
between the two electrons on the SFSs 
from interactions between original electrons. 
The transformation matrices ${\cal T}_{X\alpha\alpha'}^{(ss')}$ 
for three types of interactions, {\it i.e.}, $V_{ij}^{{\rm c}}$, 
$J_{ij}^{z}$, and $J_{ij}^{\perp}$, are obtained. 
We have examined two kinds of order-parameter mixing effects 
in such superconductors due to the strong spin-orbit coupling: 
One is the parity mixing of the spin-singlet pairs 
$\langle {\hat d}_0(\vk) \rangle$ 
and the triplet pairs $\langle {\hat d}(\vk) \rangle$, 
and the other is the interband mixing of the pairs on different SFSs, 
{\it i.e.}, $\langle {\tilde \psi}_{++}(\vk) \rangle$ 
and $\langle {\tilde \psi}_{--}(\vk) \rangle$, 
due to interband pair hopping.

First, we have examined the equal-band limit, 
where $\lambda_{s}$, $\lambda'_{s}$, 
and $\omega_{\rm c}^{(s)}$ do not depend on the band index $s = \pm$. 
Note that this limit does not imply the absence of spin-orbit coupling, 
because the split of the Fermi surfaces is taken into account 
by setting $\langle {\tilde \psi}_{\pm \mp}(\vk) \rangle = 0$. 
In this limit, 
since the parity mixing effect is suppressed, 
a pure spin-singlet state or a pure spin-triplet state occurs, 
while the interband mixing effect becomes most efficient. 
Second, we have examined the single-band limit, where $\rho_{-}(0) = 0$. 
In this limit, 
the amplitudes of the spin-singlet and triplet components coincide, 
in contrast to the equal-band limit. 
When one of the SFSs disappears, the interband mixing effect disappears, 
while the singlet-triplet mixing effect becomes most efficient. 
Between these two limits, the drastic changes explained below take place.

It is found that interband interactions 
could enhance the superconducting transition temperature $\Tc$ markedly, 
even if they are very small, 
as demonstrated in Fig.~\ref{fig:lomegadep}. 
For example, 
when $\lambda_{+} = 0.2$, $\lambda_{-} = 0.1$, 
and $\omega_c^{(-)}/\omega_c^{(+)} = 2$, 
the $\Tc$'s of independent bands ($\lambda'_{+} = \lambda'_{-} = 0$) 
are estimated as 
$T_{c0}^{(+)}/1.13 \omega_{\rm c}^{(+)} \approx 0.00764$ 
and $T_{c0}^{(-)} \approx 0.0135 \times T_{c0}^{(+)} \ll T_{c0}^{(+)}$. 
In this case, 
small interband interactions, such as $\lambda'_{+} = \lambda'_{-} = 0.1$, 
enhance the transition temperature up to 
$\Tc \approx 3.96 \times T_{c0}^{(+)} \gg T_{c0}^{(+)}$.

This effect may partly explain 
the large difference between the $\Tc$'s observed 
in ${\rm Li_2Pd_3B}$ and ${\rm Li_2Pt_3B}$ (7 and 2.7~K, respectively). 
In the latter compound, 
some of the Fermi surfaces lose 
their partners owing to the stronger spin-orbit coupling~\cite{Shis11}, 
and they do not benefit from the interband mixing effect. 
This explanation is the case 
if Fermi surfaces without partners dominate the superconductivity 
in the latter compound. 
Their contributions to the density of states are large, 
according to the first-principles calculation 
by Shishidou and Oguchi~\cite{Shis11}.

In addition, 
we have examined the effect of the difference between the two effective 
cutoff energies $\omega_c^{(+)}$ and $\omega_c^{(-)}$. 
The difference can be large in interactions mediated by 
spin and charge fluctuations, 
because each pair of Fermi surfaces has a different nesting condition, 
which is sensitive to the shape of the Fermi surfaces. 
It is found that the transition temperature strongly depends on 
the ratio $\omega_c^{(-)}/\omega_c^{(+)}$ as shown 
in Fig.~\ref{fig:lomegadep}.

Next, we have examined models with spherically symmetric Fermi surfaces 
and ${\hat \vg}(\vk) = {\hat \vk}$ as an example. 
The resultant phase diagrams drastically change 
when one of the SFSs disappears. 
In particular, we have derived areas where the transition 
from the full-gap state to the line-node state (FLT) occurs. 
The FLT occurs in many cases; however, analyzing the phase diagrams, 
it is found that it occurs in a wider realistic parameter region for the CI, 
while in rather narrower unrealistic parameter regions 
for the SIs. 
Therefore, the CI would be the dominant pairing interaction 
in many of the systems in which the FLT is observed. 
When the \mbox{s-wave} interaction is strongly repulsive, 
for example, owing to the strong on-site (screened) Coulomb repulsion, 
and the \mbox{p-wave} and \mbox{d-wave} interactions are attractive, 
we obtain a large region where the FLT occurs 
(see Fig.~\ref{fig:transition_charge}).

Therefore, the inexistence of the partners in some of the SFSs 
in ${\rm Li_2Pt_3B}$, 
which has been found by Shishidou and Oguchi~\cite{Shis11}, 
may play an essential role in the differences of the superconductivity 
in ${\rm Li_2Pt_3B}$ from that in ${\rm Li_2Pd_3B}$. 
If the FLT occurs and 
the present scenario is the case in these compounds, 
it is most likely that the full-gap state in ${\rm Li_2Pd_3B}$ 
and the line-node state in ${\rm Li_2Pt_3B}$ are 
an \mbox{s-wave} nearly spin-triplet state and 
a \mbox{d-wave} state that has both spin-singlet and triplet components 
of comparable weights, respectively, 
which are induced by the CI. 
On the other hand, 
if the full-gap state also occurs in ${\rm Li_2Pt_3B}$~\cite{Haf09}, 
both states are of \mbox{s-wave} pairing, 
which are a nearly spin-singlet state in ${\rm Li_2Pd_3B}$ and 
a singlet-triplet mixed state of comparable weights 
in ${\rm Li_2Pt_3B}$.

\begin{table}[hbtp]
\vspace{2ex} 
\vspace{2ex} 
\begin{center}
\caption{
Anomalous spin-triplet states and properties of interactions. 
The signs are those for the major parts of the areas in which 
anomalous spin-triplet states occur. 
See corresponding phase diagrams. 
The $+$ sign means that the interaction is 
attractive between original electrons. 
}
\label{table:AS}
\begin{tabular}{c}
\\
\begin{tabular}{l|c|c|c|l}
Type of ~ & \multicolumn{3}{c|}{ Sign of $\lambda_{l}$ } 
                      & ~ Anomalous  \\
\cline{2-4} 
\raisebox{1.0ex}{interaction} 
             & ~s~ & ~p~ & ~d~ & 
\raisebox{1.0ex}{~ triplet state} \\
\hline 
Charge          & $\pm$ & $+$  & $\pm$ & ~ s-wave \\
Ising spin      & $-$   & $+$  & $-$   & ~ d-wave, $Y_{20}$ \\
                & $-$   & $-$  & $-$   & ~ d-wave, $Y_{2,\pm 2}$ \\
Planar spin     & $-$   & $-$  & $-$   & ~ s-wave \\ 
Isotropic spin~ & $-$   & $-$  & $-$   & ~ s-wave \\
\end{tabular}
\end{tabular}
\end{center}
\end{table}

It is found that the magnetically mediated pairing interaction can 
induce the \mbox{d-wave} spin-triplet states 
as well as the \mbox{s-wave} spin-triplet state. 
We summarize the relation between the anomalous spin-triplet states 
and the properties of the interaction in Table~\ref{table:AS}. 
The magnetic anisotropy of the Ising-type interaction plays 
an essential role in the occurrence of the \mbox{d-wave} spin-triplet states, 
as summarized in Table~\ref{table:AS}. 
The terms $2\lambda_{z1}/15$ 
  in $\lambda_{20}$ and $\lambda'_{20}$ 
and the terms $- 4\lambda_{z1}/5$ 
   in $\lambda_{2,\pm 2}$ and $\lambda'_{2,\pm 2}$ 
in \eq.{eq:lambda_z} 
induce the \mbox{d-wave} spin-triplet states, 
when $\lambda_{z1} > 0$ and $\lambda_{z1} < 0$, respectively. 
Furthermore, when the spin-spin interaction is planar or isotropic, 
a repulsive \mbox{p-wave} interaction can induce 
the \mbox{s-wave} spin-triplet state, 
if the even-parity interactions are repulsive or weak.

Lastly, we discuss the experimental result of the Knight shift 
in ${\rm Li_2Pt_3B}$~\cite{Nis07}, 
which exhibits a flat temperature dependence. 
As we mentioned in \S~\ref{sec:introduction}, 
it seems that conventional theory could not explain this result. 
If we assume that the spin susceptibility remains unchanged across $\Tc$, 
we obtain 
$\langle {\hat d}_z(\vk) \rangle = \langle {\hat d}_0(\vk) \rangle = 0$ 
for the majority of $\vk$'s. 
Since $\langle {\hat d}_z(\vk) \rangle 
= \langle {\hat d}(\vk) \rangle {\hat g}_z(\vk)$, 
we obtain $\langle {\hat d}(\vk) \rangle = 0$ 
or ${\hat g}_z(\vk) = \cos {\bar \theta}_{\vk} = 0$. 
Therefore, 
this leads to a contradiction that 
all the superconducting gap functions vanish as 
$\langle {\hat \vd}(\vk) \rangle 
= \langle {\hat d}(\vk) \rangle \, {\hat \vg}(\vk) = {\vector 0}$ 
and $\langle d_0(\vk) \rangle = 0$, 
unless the superconductivity occurs mainly on parts of Fermi surfaces 
in which ${\bar \theta}_{\vk} = \pi/2$ is satisfied. 
However, in ${\rm Li_2Pt_3B}$, since the sample was powder, 
the angles between the magnetic field and the crystal axes would have 
distributed randomly. 
One of the possible explanations for this is that the states of the sample, 
such as the gap function of the superconductivity 
and the orientations of the powders, 
are considerably affected by the magnetic field applied in the measurement. 
A theoretical interpretation of the behavior of the Knight shift 
remains for future studies.

In conclusion, 
the superconductivity in noncentrosymmetric system drastically changes 
when one of the SFSs vanishes as the spin-orbit coupling increases. 
For example, the gap structures, transition temperatures, 
and phase diagrams are quite different 
depending on whether both SFSs exist. 
In particular, under some conditions, 
the FLT occurs when one of the SFSs disappears. 
The area of the FLT in the phase diagram is largest 
when the pairing interaction is the CI 
and the condition 
$\lambda_{{\rm c}1} > \lambda_{{\rm c}2} > 0 > \lambda_{{\rm c}0}$ 
is satisfied. 
The latter condition seems realistic in real materials 
if we assume the CI. 
Therefore, within the present theory, 
it is most likely that the CI is the dominant pairing interaction 
in systems in which an FLT occurs, 
although possibilities of magnetically induced pairing interactions 
are not excluded. 
Anomalous superconducting states, 
such as the \mbox{s-wave} and \mbox{d-wave} spin-triplet states, 
are induced by an attractive or repulsive \mbox{p-wave} spin-triplet interaction 
in the presence of interband spin-triplet-pair hopping interactions, 
which are active only when both SFSs exist. 
These behaviors are sensitive to the type of pairing interaction.

\begin{center}
{\bf ACKNOWLEDGMENTS}
\end{center}


We wish to thank T.~Shishidou for useful discussions and information 
on their results of the first-principles calculation 
in ${\rm Li_2Pd_3B}$ and ${\rm Li_2Pt_3B}$. 
We are very grateful to H.~Tou for useful discussions 
and information on practice and analysis in NMR experiments. 




\end{document}